\documentclass[galaxies,article,submit,pdftex,moreauthors]{Definitions/mdpi} 

\firstpage{1} 
\pubvolume{1}
\issuenum{1}
\articlenumber{0}
\pubyear{2024}
\copyrightyear{2024}
\datereceived{ } 
\daterevised{ } 
\dateaccepted{ } 
\datepublished{ } 
\hreflink{https://doi.org/} 

\usepackage{bm}
\usepackage{MnSymbol} 
\usepackage{adjustbox}

\graphicspath{{Definitions/imgs/}}

\Title{Numerical study of bar suppression in galaxy models due to disc heating}

\TitleCitation{Numerical study of bar suppression in galaxy models due to disc heating}

\Author{Alejandro López Gómez $^{1}$*, Ruslan F. Gabbasov $^{2}$ and Isaura Luisa Fuentes-Carrera $^{1}$}

\AuthorNames{Alejandro López Gómez, Ruslan F. Gabbasov and Isaura Luisa Fuentes-Carrera}

\AuthorCitation{López Gómez, A.; Gabbasov, Ruslan F.; Fuentes-Carrera, I. L.}

\address{%
$^{1}$ \quad Escuela Superior de F\'isica y Matem\'aticas, Instituto Polit\'ecnico Nacional, U.P. Adolfo L\'opez Mateos, \hspace{0.2cm}edificio 9, Zacatenco, 07730 Ciudad de M\'exico, M\'exico\\
$^{2}$ \quad Universidad Aut\'onoma Metropolitana, Alcald\'ia Azcapotzalco, 02200 Ciudad de M\'exico, M\'exico}

\corres{Correspondence: alelopez@ipn.mx}

\abstract{The process of bar formation, evolution and destruction is still a controversial topic regarding galaxy dynamics. Numerical simulations show that these phenomena strongly depend on physical and numerical parameters. In this work, we study the combined influence of the softening parameter, $\epsilon$ and disc mass fraction, $m_{\mathrm{d}}$ on the formation and evolution of bars in isolated disc-halo models via $N$-body simulations with different particle resolutions. Previous studies indicate that the bar strength depends on $m_{\mathrm{d}}$ as $\propto m_{\mathrm{d}}^{-1}$, which is seen as a delay in bar formation. However, the distorsion parameter, $\eta$, which measures the bar's momentum through time, shows that an increase in $m_{\mathrm{d}}$ does not always induce a delay in bar formation. This suggests that $\epsilon$ interact to either enhance or weaken the bar. Moreover, numerical heating dominates in models with small softening values, creating highly accelerated particles at the centre of discs, regardless of $m_{\mathrm{d}}$ or resolution. These enhanced particle accelerations produce chaotic orbits for $\epsilon \leq 5$\,pc, resulting in bar suppression due to collisional dynamics in the centre. In our high resolution models ($N \approx 10^{7}$), small softening values are incapable of reproducing the bar instability. The role of disc mass is as follows: increasing $m_{\mathrm{d}}$ for moderate $\epsilon$ ($\geq 10$\,pc) reduces the amount of drift in the acceleration profile, without affecting the bar's behaviour. Models with lower $m_{\mathrm{d}}$ values coupled with small softening values, have an excess of highly accelerated particles, introducing unwanted effects into otherwise reliable simulations. Finally, we show that the evolution of the disc's vertical acceleration profile is a reliable indicator of numerical heating introduced by $\epsilon$ and the bar.}


\begin{document}

\section{Introduction} \label{sec:introduction}

\begin{figure}[t]
\centering
\includegraphics[scale=0.65]{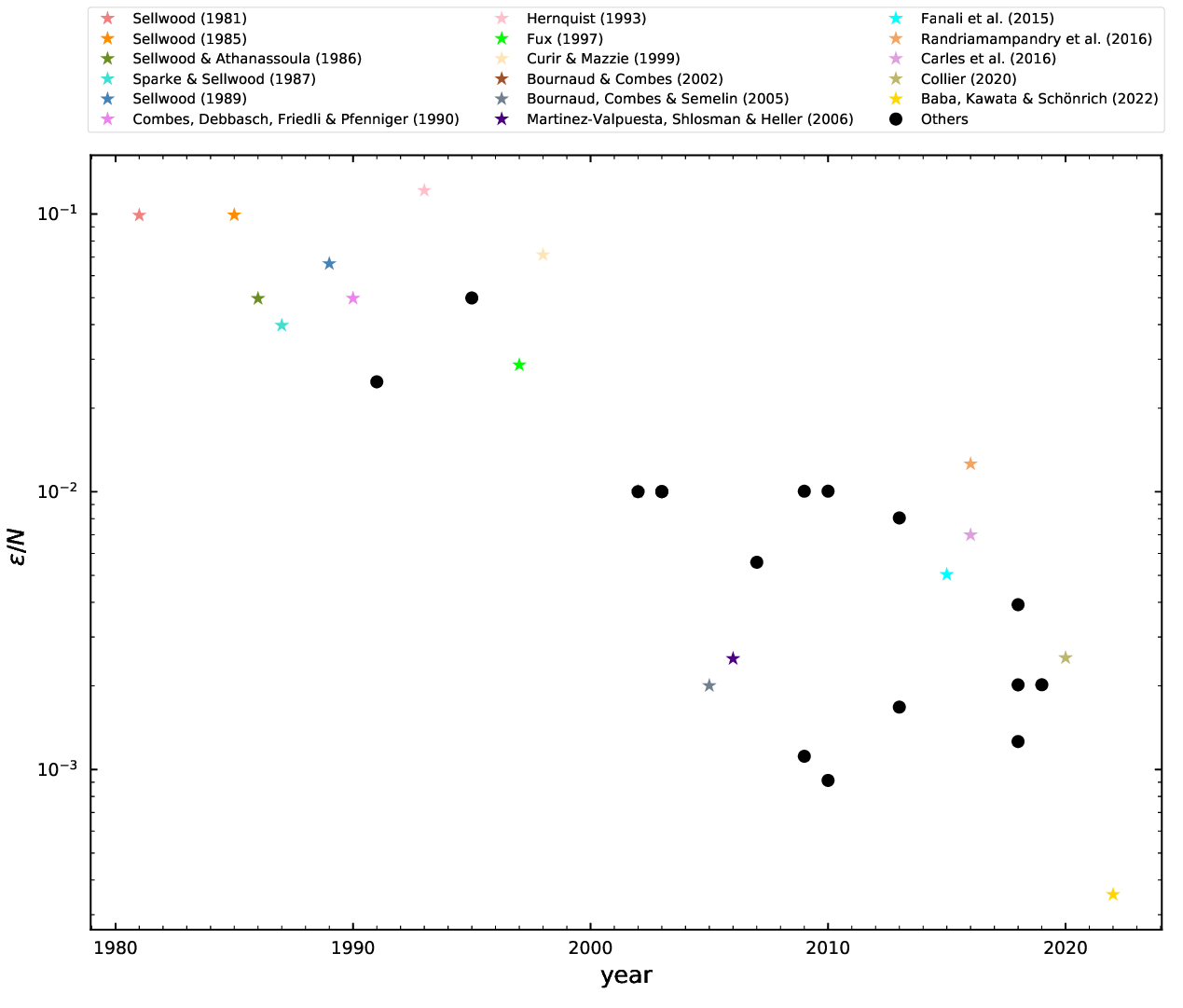}
\caption{Values of $\epsilon$ used throughout time for galactic-scale $N$-body simulations normalized by $N$. Annotations for some of the paper's authors are located at the right of the plot and are denoted in our plot by colored stars; the rest of simulations are plotted with a filled black circle. 
The softenings in this figure are in parsecs (pc).}
\label{fig:softening-time}
\end{figure}


Bar formation, evolution and destruction are all long-standing subjects in Astrophysics. Extensive efforts trying to understand bar phenomena include studies of theoretical \citep{Lindblad1963,LinShu1964,Toomre1964,Lynden-BellKalnajs1972,Toomre1981}, numerical \citep{Sellwood1981,DubinskiKuijken1995,AthanassoulaMisiriotis2002,YurinSpringel2015}, and observational \citep{Kormendy1981,Elmegreenetal1990,SellwoodWilkinson1993,KormendyKennicutt2004,Mastersetal2010} nature. Numerical simulations approach bar phenomenology from three distinctive fronts: 
gravitational-only \citep{DebattistaSellwood1998,ValenzuelaKlypin2003,PetersenWeinbergKatz2016,Valencia-Enriquezetal2017}, hydrodynamical \citep[including collisionless particles,][]{Chilingarianetal2010,HwangParkChoi2013,Dobbsetal2018,SellwoodShenLi2019}, and cosmological \citep{ScannapiecoAthan2012,Schayeetal2015,Crainetal2015,Algorryetal2017}, 

Studies of structure (and substructure) formation in isolated galaxies in pure $N$-body simulations have expanded our understanding of bar dynamics \citep{OstrikerPeebles1973,Hohl1976,Hohl1978,AthanassoulaMisiriotis2002,SahaNaab2013}. Despite of lacking gas physics and all the effects that come with it (star formation, feedback, etc.), $N$-body experiments are able to reproduce a lot of fundamental properties involving bar-like phenomena~\citep{LittleCarlberg1991,HarsoulaKalapo2009,XingchenShlosmanetal2023}. For example, \cite{DebattistaSellwood2000} performed a series of $N$-body simulations composed of a Dark matter (DM) halo and a stellar disc, covering different types of halos (rotating, non-rotating, etc.) and values for Toomre's criterion, $Q$, ranging from 0.05 to 1.5; with $N = 2\times 10^{5}$ particles and a resolution of $\epsilon \simeq 0.7\,$kpc for their Plummer softening length. They showed that bars tend to slowdown, independently of the DM halo inner profile or its velocity distribution, by dynamical friction \citep{Weinberg1985} between bar and DM halo. Another important outcome of \cite{DebattistaSellwood2000} is that bars persevere after periods of extreme friction, even if the slowdown is very dramatic, which means that bars are robust objects; something that was doubtful in previous works \citep[e.g.][]{Kormendy1979}. Later on, \cite{ValenzuelaKlypin2003} evolved similar disc-halo pairs but vastly improving the number of particles, $N = 3.55\times 10^{6}$ and numerical resolution with $\epsilon \simeq 0.22\,$kpc (values corresponding to their model $A_{1}$), and found results that contradicted the conclusions of \cite{DebattistaSellwood2000}. They demonstrated that the low particle number and choice of $Q$ used by \citep{DebattistaSellwood2000} caused a numerical artifact that led to low bar pattern speeds, a fact also verified by \cite{Athanassoula2003}. \cite{ValenzuelaKlypin2003} also found differences in the lengths of their bars, especially compared to those reported by \cite{AthanassoulaMisiriotis2002}. While these differences are relevant, both works agree on the fact that a recently formed bar tends to grow and does not disappear spontaneously. 

Angular momentum exchange between halo and disc also seems to play a prominent role in bar growth and stabilization \citep{WeinbergKatz2007,Dubinskietal2009,Sahaetal2012,Athanassoula2013}. Bars form due to resonant orbits induced by sudden fluctuations in the disc's density distribution~\citep{Athanassoula2013}. These resonant orbits promote angular momentum transfer between the disc and an external component that surrounds it (usually a DM halo). If the density perturbation has the right wavelength, the transfer of momentum occurs in a feedback loop that only benefits bar growth. Now, if we were able to shutdown the channels\footnote{In this case, resonant orbits, mostly of the $x_{1}$ family.} that allow the transfer of angular momentum, we could, in principal, \emph{destroy} a bar \citep{HasanNorman1990,PfennigerNorman1990,HozumiHernquist2005}. \cite{Kormendy2013} calls it bar ``suicide''. In short, to destroy the orbits that comprise a bar it is necessary to gather a huge amount of baryonic matter at its centre \citep[${\sim}10 \%$ of the total disc mass,][]{Athanassoulaetal2005}. In order to investigate the existence of unbarred galaxies, \cite{SahaElmegreen2018} constructed galaxy models including stellar disc, DM halo and spherical bulge. They showed that adding a central bulge component sufficiently dense (denser than the disc at the bulge half-mass radii) could prevent the formation of a bar instability. None of their models had a gas component but, as stressed by these authors, a dissipative element would only weaken the bar \citep[see, for example][]{Berentzenetal1998}. 


The paramount achievements listed above have one glaring flaw: the way numerical parameters are chosen. We have already mentioned some of the parameters that usually relate to bar dynamics, such as $N$ or $\epsilon$. In addition, gas dynamics and halo physical properties like the spin parameter $\lambda$, concentration $c$ and virial velocity $v_{200}$ are known to also affect the formation and growth of bars~\citep{FoyleCourteauThacker2008,Collier2020}.
Nonetheless, here we constrain our study to $\epsilon$, $N$ and the disc mass fraction, $m_{\mathrm{d}}$. These parameters are involved in how bars are formed and, although the influence of $m_{\mathrm{d}}$ has already been established by $N$-body simulations of disc galaxies~\citep{ValenzuelaKlypin2003,Athanassoula2003}, it is not well-known how $\epsilon$ truly affects them and whether or not it interacts with other parameters. In $N$-body simulations it is necessary to \textit{soften} the forces between particles to, on one hand, stay in collisionless regime and avoid \textit{relaxation} (too small softening) and, on the other, to avoid excessive smoothing \textit{damping} (too large softening). Moreover, simulations must include a softening kernel that maps the gravitational potential to correctly distribute it along the radial extension of each component. The most common is the Plummer softening. However, it is known that the Plummer softening behaves quite poorly when resolving the frequencies that enhance bar modes~\citep{deRijckeFouvryDehnen2019}. In fact, any kind of gravitational softening tends to lower the growth of $m=2$ modes, although kernels such as the cubic spline are able to resolve them well enough~\citep{deRijckeFouvryDehnen2019}.

Balance amid relaxation and damping is key to produce realistic $N$-body simulations. However, the process to achieve such balance, and therefore to set a proper softening length for numerical models, is difficult and subject to a large number of uncertities. Formerly, scientists adapted $\epsilon$ due to its relationship with $N$. For a given $N$, the value for $\epsilon$ is chosen such that the error between sampling the density distribution and shot noise are minimized~\citep{Merritt1996,Athanassoulaetal2000,Dehnen2001,RodionovSotnikova2005}.

Another approach was given by \cite{Romeo1994}. He studied the effect of softening for one-component (stellar disc) and two-component (stellar and gaseous discs) two-dimensional $N$-body simulations, stressing that (i) for one-component systems, high softening values ($\epsilon \gtrsim 0.36$\,kpc) lack physical consistency because they either have no physical equivalent (like the scale height for small softening values) or tend to quell most instabilities artificially;
(ii) for both systems, when $0.20$\,kpc $\lesssim \epsilon \lesssim 0.36$\,kpc, discs are stable but softening often induces artificial quenching on spiral modes, depending on the wavelength of the perturbation and
(iii) models with Newtonian gravity, i.e. $\epsilon \to 0$, are unstable for a wide range of perturbations, especially those comparable with the characteristic wavelength of the stellar or gaseous components.
The above assertions are tightly constrained by the ``temperature'', $Q$, of the discs. Note that Romeo's simulations \cite{Romeo1994} do not account for the stabilizing effect of a DM halo.

The adopted choices found in these studies assumed that the density distribution is static and the particles are of equal mass; this is not true either for simulations of cosmological or galactic scale. Specifically, $N$-body galaxies are, in general, multi-component and, at times, multi-phase systems, meaning that the particle distribution is not constant. 
Codes like \textsc{gadget} \citep{Springel2005} coped with this by adding softening values for each component. The components can have different smoothing lengths and different masses by demanding that the minimum gravitational potential has the same value for all particles.
This allows to simulate galaxies with much less halo particles without significant increase in the relaxation.

More recently, \cite{IannuzziAthanassoula2013} simulated collisionless barred galaxies with the purpose of evaluating the influence of softening in bar formation, finding virtually no difference between models using an adaptative algorithm designed to adjust $\epsilon$ when local particle concentration changes and models with a fixed value of $\epsilon$ for the entire run; result which is, at the very least, unexpected. 

In Fig.~\ref{fig:softening-time}, we summarize the ranges of softening values normalized by the number of particles that have been used in galactic-scale $N$-body simulations\footnote{We do not distinguish between Plummer or kernel based softenings. Some of these simulations may also include hydrodynamical effects.}. There has not been considerable change in softening ranges after the year 2000 despite almost exponential growth of $N$.
One would expect a decrease of $\epsilon$ as $N$ increases in time by following the notion that for higher $N$ the amount of errors in force calculation minimizes when $\epsilon$ decreases~\citep{Athanassoulaetal2000,Zhan2006}. Instead, some authors use other recipes to adjust their softening values, such as stability analysis~\citep{Romeo1998} or the interparticle separation~\citep{Gabbasov2006}, which may explain the scatter in Fig~\ref{fig:softening-time}. On the other hand,
the spread of $\epsilon$ in the last two decades may be attributed to the fact that some authors reported values for Plummer softening or its equivalent for other softening kernels.

Most simulations are constrained to $\epsilon < 0.2$\,kpc and most simulations after the year 2000 are constrained to $\epsilon < 0.1$\,kpc. All of the simulations in Fig.~\ref{fig:softening-time} contain more than one component, meaning that they may use multiple softening lengths, one for each component. In such cases, we usually take the lowest value for $\epsilon$, regardless of the type of component it is used on, but we restrain the set to non-gaseous components.

\subsection{Paper overview} \label{sec:paper-overview}

In order to establish the role of softening, disc mass and number of particles on bar formation and evolution, we made a series of simulations and measured the disc and bar properties, such as the tangential average velocity, velocity dispersions, radial accelerations and bar strength throughout 12 Gyrs.

Our paper is organized as follows: section \ref{sec:methods} describes the initial conditions of $N$-body disc galaxies, details of simulations and their analysis. Section~\ref{sec:analysis} discusses and applies the parameters used to assess their disc stability. In section~\ref{sec:results} we revise and discuss the results obtained throughout this paper, including subsections describing the evolution of the models (sections \ref{sec:bar-evol-velocity-curves} and \ref{sec:evol_eta}) and their numerical consistency (section \ref{sec:particle-resolution}). Finally, in section \ref{sec:conclusions}, we give conclusive notes about the results and implications of this work.

\section{Methods} \label{sec:methods}

We describe the methods and techniques to achieve the simulations presented in this work. We follow the usual methodology, i.e. we propose an array of models that comply with our purposes, then the models are evolved and analysed to obtain a qualitative (or quantitative) relation between the numerical parameters. From here on, the softening values are given in units of kpc, unless indicated otherwise.

\begin{figure}
\centering
\includegraphics[scale=0.50]{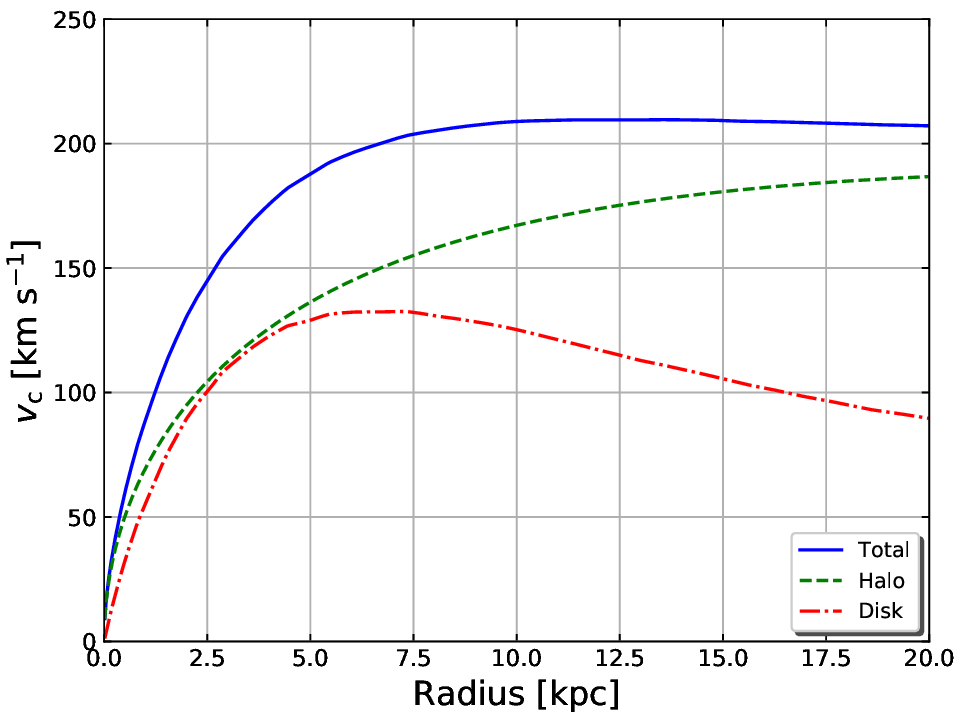}
\caption{Circular velocities for the DM halo (green dashed line), stellar disc (red dashed-dotted line) and the contribution of both (blue solid line), given by equations~(\ref{eq:rhoh}) and~(\ref{eq:rhod}). These curves correspond to model SGS2 in Table~\ref{tab:t1}.}
\label{fig:rot_curve_init}
\end{figure}

\subsection{Choice of parameters} \label{sec:choice-parameters}

There are several studies that attempt to relate the numerical properties of $N$-body simulations to their physical behavior~\citep{Gabbasov2006,IannuzziAthanassoula2013}. 
In this work, we seek to constrain the range of two parameters: the disc mass fraction, $m_{\mathrm{d}}$ and the softening length, $\epsilon$. Now, for this study to be relevant in regards to reproducing actual bar phenomena, we concentrate on those sets of values that have the most influence on disc dynamics~\citep{Sellwood2013}. It has also been shown that bar dynamics in an isolated galaxy is dominated by the disc and DM halo components \citep{Sellwood2013}, and although structures like a classical bulge or a thick external gas disc indeed affect the secular evolution of bars, their role is secondary in comparison to both disc and halo.

We break down how to select the parametrical window for both $m_{\mathrm{d}}$ and $\epsilon$ down below. It is important to mention that our intent is not to find an ``optimal'' range of values for $\epsilon$~\citep[contrary to, for example,][]{Merritt1996}. Our main objective is to assess the influence of both $m_{\mathrm{d}}$ and $\epsilon$ on bar formation and evolution, and whether these parameters interact to benefit or hinder its development~\citep[see][]{Villa-VargasShlosmanHeller2010}.

The process to choose a proper array of values for $\epsilon$ is, in some regards, arbitrary. Softening is a particularly difficult parameter to constrain, especially for simulations of galactic scale, because there is no direct analogue with observational parameters. Despite numerous attempts to find an appropriate softening value \citep[e.g.][]{Romeo1998,Dehnen2001,Hobbs2016}, there is still no way to determine an ideal numerical value for $\epsilon$. The fact that $N$-body simulations are subjected to a large number of free parameters is to blame. Numerical parameters such as $\epsilon$ appear to be unrelated to other physical parameters, despite having a tremendous impact on the dynamical properties of $N$-body galaxies. However, parameters like $m_{\mathrm{d}}$ and $\epsilon$ are clearly related to each other through the equations of motion, which in turn are involved in force calculation. This is because the force for each particle within the density profile is proportional to the mass of the particle, $m$, and inversely proportional to $\epsilon^2$, and such mass changes due to $m_\mathrm{d}$ as $m=m_{\mathrm{d}} M_{200}/N_{\mathrm{d}}$, for the disc, and $m=M_{200}(1-m_{\mathrm{d}})/N_{\mathrm{h}}$, for the halo. This means that constraining $\epsilon$ should not be reduced to studying its relation to the particle resolution, but also include its interaction with other relevant quantities.

Another important factor is the softening kernel introduced in the modification of the Newtonian gravity. \textsc{gadget-2} uses a cubic spline kernel \citep{MonaghanLattanzio1985}, which is an improvement from the commonly used Plummer softening kernel because the spline kernel is exactly Newtonian at several softening lengths. For historical reasons, the spline softening length $h$ is related to the Plummer softening length as $h = 2.8\epsilon$. 

So, how to pick the right softening length for our models if there is no ``correct'' choice yet? The simplest approach is to choose values from some of the studies already conducted by other authors \citep[e.g.][]{Springel2001,Gabbasov2006,IannuzziAthanassoula2013} that bring the best results, and then expand our study to limits where calculated forces could become artificial. For example, \cite{IannuzziAthanassoula2013} used $\epsilon = 0.006$ as lower limit in their adaptative softening code. They obtain this values by finding the number of neighbors in the density field for a given particle and mapping the result to the softening kernel (in this case, the cubic spline); the more distance there is between neighbors, the larger the softening length is. Something that stands out from these simulations is that the softening length distribution does not appear to change throughout their simulations, despite them clearly displaying a bar at the end (${\sim}10$\,Gyrs). They also demonstrate that changes in the fixed softening value do not affect the behaviour of the simulations\footnote{\cite{IannuzziAthanassoula2013} use a fiducial value of $\epsilon = 0.05$ in their fixed-softening simulations. They also ran simulations with $\epsilon = 0.025, 0.1$}.
Large softening values ($\epsilon > 0.1$) normally bring spurious outcomes \citep{Romeo1994,Romeo1997,Romeo1998}, but there is no agreement to what is the upper limit for the appearance of serious discretization effects in force calculation. We propose to use an upper limit of $\epsilon = 0.1$. We select $\epsilon = 0.001$ as our lower limit. We justify the latter selection given that several numerical experiments already use gravitational softening lengths close to $0.001$~\citep[e.g.][]{Dubinskietal2009,Collier2020}. Another reason is that we are interested in studying the behavior of $N$-body discs when particles come close to being collisional. We use the same fixed softening length for both disc and DM halo. This is because our initial conditions use the same softening value to perform force calculation (see section~\ref{sec:ics}). It is worth noting that one may use different softening lengths for multicomponent systems~\citep[e.g.][]{McmillanDehnen2007}, choosing its values so that the force between all particles is the same regardless of mass. We consider such setting in appendix~\ref{sec:apen1} by comparing the evolution of three models: a low resolution model ($N\sim 10^6$), a high resolution model ($N \sim 3\times 10^6$) and a low resolution model where the halo softening length, $\epsilon_{\mathrm{h}}$, is given by

\begin{equation}
\epsilon_{\mathrm{h}} = \frac{m_{\mathrm{h}}}{m_{\mathrm{d}}} \epsilon_{\mathrm{d}},
\label{ec:same_softening}
\end{equation}

where $m_{\mathrm{d}}$ is the mass of a disc particle, $m_{\mathrm{h}}$ is the mass of a halo particle and $\epsilon_{\mathrm{d}}$ is the disc softening.

The amount of disc mass assigned to $N$-body simulations is roughly constrained by cosmological numerical experiments between $0.02M$ to $0.2M$~\citep{MMW1998}, where $M$ is the total mass of the galaxy. For example, \cite{ZhouZhuWangFeng2020} studied over two thousand disc galaxies of the Illustris-1 \citep{Vogelsbergeretal2014} and TNG100 \citep{Nelsonetal2018} sets of cosmological simulations. They find that stellar discs have bars in ${\sim}30{\%}$ of them when stellar masses are $M_{\star} \gtrsim 10^{11.25} M_{\odot}$ on the Illustris-1 catalogue and that, for the TNG100 catalogue, there is a fraction of ${\sim}50{\%}$ of barred discs when stellar masses are $M_{\star} = 10^{10.66-11.25} M_{\odot}$. In both catalogues, the barred discs fraction decreases when the stellar mass decreases. Galaxies with lower stellar masses ($3.3 \times 10^{10} M_{\odot} < M_{\star} < 8.3 \times 10^{10} M_{\odot}$) rarely have bars; such results are redshift-dependent. This range can be also found via observational constraints. Using the Tully-Fisher relation \citep{TullyFisher1977} for a sample of 81 disc galaxies and taking into account adiabatic contraction, \cite{GnedinWeinbergetal2007} find the average disc mass fraction, $\tilde{m}_{\mathrm{d}}$, that best fits each object. All the fitted galaxies fall within the range $0.02 \leq \tilde{m}_{\mathrm{d}} \leq 0.1$, which is in agreement with other studies of the same nature~\citep[e.g.][]{GovernatoWillman2007,Wu2018}. Therefore, assuming a lower limit of $3.3 \times 10^{10} M_{\odot}$ and an upper limit of $10^{11.25} M_{\odot}$ for the stellar mass of galaxies, and using a virial velocity for the DM halo\footnote{This value roughly corresponds to the mass of Milky Way-size galaxies, given that $M_{200} \propto v_{200}^3$.} of $v_{200} = 160$\,km\,s$^{-1}$, it is possible to use $0.033 \leq m_{\mathrm{d}} \leq 0.17$ as a plausible range for the disc mass fraction, which is not far from the values estimated by theoretical considerations.

Taking the above into account, we now impose a range of values for $m_{\mathrm{d}}$. Since we only have a disc and a DM halo, the total mass of our systems is given by $M = M_{\mathrm{d}} + M_{\mathrm{h}}$. Disc and DM halo masses are $M_{\mathrm{d}} = m_{\mathrm{d}} M$ and $M_{\mathrm{h}} = (1 - m_{\mathrm{d}}) M$, respectively. As mentioned before, realistic discs form when the disc mass fraction falls within the range $0.02 \leq m_{\mathrm{d}} \leq 0.2$; this range is in agreement with the ranges given by cosmological simulations and observations. However, not all discs in this range are susceptible to bar formations. Along with the models presented in this work, we also performed simulations with a wider range of values for the disc mass fraction ($0.02 \leq m_{\mathrm{d}} \leq 0.2$)\footnote{We ran a couple of experiments using $m_{\mathrm{d}} = \{0.02,0.3\}$ that confirm our results.}. These experiments indicate that bars form roughly within the range $0.02 \leq m_{\mathrm{d}} \leq 0.08$, and most of the bars with $m_\mathrm{d}$ close to $0.08$ are considerably weak. So, with this information in mind, we chose $0.035 \leq m_{\mathrm{d}} \leq 0.05$, interval that allows stellar discs to be stable in a global sense but at the same time still susceptible to local instabilities, like bars. 

The configuration of the galaxy models is completed by specifying the \textit{spin} parameter $\lambda$, and the fraction of angular momentum of the disc with respect to the halo's angular momentum, $j_{\mathrm{d}}$.
In order to avoid any unphysical values for scale length and velocity structure on the disc particles, we adopt the fiducial convention of $m_{\mathrm{d}} = \lambda = j_{\mathrm{d}}$~\citep{MMW1998,SpringelWhite1999,Springel2001}. In this way, we ensure that disc and halo are embedded properly, at least according to the standard galaxy formation theory. This configuration also means that $m_{\mathrm{d}}/\lambda = 1$, which does not necessarily benefit the development of bars in our discs, but leaves the door open for other parameters to disrupt their stability~\citep{FoyleCourteauThacker2008}.

\subsection{Initial conditions} \label{sec:ics}

To generate proper initial conditions (ICs), we use \textsc{galic} \citep{YurinSpringel2014}, a code created with the specific purpose of generating disc galaxies that, ideally, are in perfect equilibrium. \textsc{galic} applies Rodionov's method to find iterations for Poisson's equation that comply with conditions suited to a certain equilibrium state using the so-called \textit{merit function} \citep{Rodionov2009}. 
\textsc{galic} defines two merit functions, one to map the desired density distribution, and another for the velocity structure of said distribution. The latter is needed because it is not enough to use one merit function to approximate the density distribution of a system. In turn, one must also minimize the velocity structure so that it resembles the correct conditions set by the specified profile. This method allows \textsc{galic} to find precise values for the distribution function of well-known density profiles (see below). The iterative method in \textsc{galic} has the disadvantage of being time and resource consuming in the long run, so we limit the number of iterations to 10, which is the same number of iterations recommended by those authors.

We must also mention that \textsc{galic} calculates the forces for each component in the ICs using the same $\epsilon$ in the multipole expansion, so we must apply the same softening for all particles in our models, regardless of the species.
We use the same fixed softening length in both ICs and simulations.

\subsubsection{Dark matter halo} \label{sec:dmh}

The halo distribution is the same as in \cite{SpringelWhite1999} and is equivalent to the profile proposed by \cite{Hernquist1990}. The particle distribution inside a radius $r$ is defined by 
\begin{equation}
\rho_{\mathrm{h}}(r) = \frac{M_{\mathrm{h}}}{2 \pi} \frac{a_{\mathrm{h}}}{r(r + a_{\mathrm{h}})^{3}},
\label{eq:rhoh}
\end{equation}
where $a_{\mathrm{h}}$ and $M_{\mathrm{h}}$ are the scale factor and mass of the DM halo, respectively. This distribution can be associated with a NFW halo~\citep{NFW1996} through the concentration parameter, $c$, which is given by
\begin{equation}
a_{\mathrm{h}} = \frac{r_{200}}{c} \sqrt{2 \left[ \ln (1+c) - c/(1+c) \right]},
\label{eq:cpar}
\end{equation}
where $r_{200}$ is the virial radius of the NFW halo and is related to its virial mass, $M_{200}$. It can be assumed that $M_{200} = M_{\mathrm{h}}$ \citep{MMW1998}. In this case, the tangential velocity is essencially one of a \textit{cuspy} DM halo and is depicted in Fig.~\ref{fig:rot_curve_init}.

\subsubsection{Stellar disc} \label{sec:disc}

The disc distribution is an exponential radial profile with a $\mathrm{sech}^{2} (z)$ vertical profile, which is given by
\begin{equation}
\rho_{\mathrm{d}}(R,z) = \frac{M_{\mathrm{d}}}{4 \pi z_{0} R_{\mathrm{d}}^{2}} \mathrm{e}^{-R/R_{\mathrm{d}}} \mathrm{sech}^{2}(z/z_{0}).
\label{eq:rhod}
\end{equation}
Here, the disc scale length, $R_{\mathrm{d}}$, is calculated from the prescription given in \cite{MMW1998}, which contains a direct dependency on the disc mass fraction. The scale height $z_{0}$ is assumed constant and in all our models, $z_{0} = 0.2 R_{\mathrm{d}}$.

The model shown in Fig.~\ref{fig:rot_curve_init} resembles model D1 in \cite{YurinSpringel2014}, which was tested by the authors with $N_{\mathrm{h}} = 10^{6}$ and $N_{\mathrm{d}} = 10^{6}$, $\epsilon = 0.05$ and $m_{\mathrm{d}} = 0.035$. This model has a stability threshold of $\min(Q_{t=0}) \equiv Q_{\mathrm{min}} \approx 1$ and the authors report that it is the most susceptible of their models to experience axisymmetric perturbations\footnote{Whether or not this model developed a bar is unknown, \cite{YurinSpringel2014} only showed the first gigayear of the simulation.}.

The halo profile is symmetric and isotropic, with $\sigma_{r}/\sigma_{z} = 1$. The disc profile has a distribution function defined by $f(E,L_z)$ and its velocity structure is that of an isotropic rotator, i.e.

\begin{equation}
 \sigma_r = \sigma_z = \sigma_{\phi} = \langle v_{\phi}^2 \rangle - \langle v_{\phi} \rangle^2,   
\end{equation}

where $\langle v_{\phi}^2 \rangle$ is the second moment of the azimuthal motion and $\langle v_{\phi} \rangle$ is the streaming in the azimuthal direction. 

In addition to the very nature of the collisionless systems we analyse, the models studied here, if compared to real galaxies, should resemble early-type systems due to their lack of a gas component. 

\begin{center}
\begin{table}[h!]
\caption{Simulations with different values of $m_{\mathrm{d}}$ and $\epsilon$. \emph{Column 1}: name of model. \emph{Column 2}: disc mass fraction with respect to halo mass parameter. \emph{Column 3}: softening length parameter in kpc. \emph{Column 4}: minimum value of Toomre's parameter for the IC ($t = 0$). \emph{Column 5}: average deviation at $t = 1.8$\,Gyrs of the simulation from ICs radial dispersion. \emph{Column 6}: average deviation at $t = 1.8$\,Gyrs of the simulation from ICs tangential dispersion. \emph{Column 7}: average deviation at $t = 1.8$\,Gyrs of the simulation from ICs vertical dispersion. \emph{Column 8}: distorsion parameter once the simulation has finished (last snapshot). \emph{Column 9}: amount of steps necessary to complete the simulation. \emph{Column 10}: timestep between each step in Gyrs. \emph{Column 11}: Radial scale length of the stellar disc's IC in kpc. \emph{Column 12}: Percentage of angular momentum lost by the disc. Every simulation lasts approximately 12\,Gyrs.}
\vspace{0.2cm}
\label{tab:t1}
\begin{tabularx}{\textwidth}{S S S S S S S S S M S S}
\toprule
\textbf{Model} & $\bm{m_{\mathrm{d}}}$ & $\bm{\epsilon$} & $\bm{Q_{\mathrm{min}}}$ & $\bm{\overline{\sigma}_{r}}$ & $\bm{\overline{\sigma}_{\mathrm{t}}}$ & $\bm{\overline{\sigma}_{z}}$ & $\bm{\eta_{\mathrm{e}}}$ & \textbf{\textit{steps}} & $\bm{\Delta t}$ & $\bm{R_{\mathrm{d}}}$ & $\bm{\% \Delta L_{\mathrm{d}, \mathrm{f}}}$ \\
\multicolumn{1}{c}{(1)}         & (2)                   & (3)              & (4)                     & (5)                          & (6)                                   & (7)                          & (8)                      & (9)                                                       & (10)            & (11)     & (12)               \\ \cline{1-12}
SGS1                            & 0.035                 & 0.1              & 1.13                 & 0.386                      & 0.086                               & 0.088                      & 0.248                  & 32768                                                     & 3.66E-04        & 2.87     & 27.63                \\
SGS2                            & -                     & 0.05             & 1.13                 & 0.383                      & 0.089                               & 0.100                      & 0.229                  & 56518                                                     & 1.83E-04        & -        & 24.29             \\
SGS3                            & -                     & 0.025            & 1.11                 & 0.389                      & 0.092                               & 0.098                      & 0.217                  & 65536                                                     & 1.83E-04        & -        & 23.93            \\
SGS4                            & -                     & 0.010            & 1.07                 & 0.421                      & 0.117                               & 0.112                      & 0.217                  & 131115                                                    & 4.58E-05        & -        & 22.39            \\
SGS5                            & -                     & 0.005            & 1.07                 & 0.434                      & 0.113                               & 0.108                      & 0.126                  & 375840                                                    & 2.29E-05        & -        & 23.22             \\
SGS6                            & -                     & 0.001            & 1.08                 & 0.489                      & 0.177                               & 0.165                      & 0.008                  & 3433757                                                   & 1.43E-06        & -        & 6.23            \\ \cline{1-12}
SGS7                            & 0.038                 & 0.1              & 1.14                 & 0.335                      & 0.054                               & 0.068                      & 0.234                  & 32768                                                     & 3.66E-04        & 3.08     & 23.26             \\
SGS8                            & -                     & 0.05             & 1.14                 & 0.358                      & 0.066                               & 0.065                      & 0.213                  & 53683                                                     & 1.83E-04        & -        & 23.03            \\
SGS9                            & -                     & 0.025            & 1.13                 & 0.373                      & 0.078                               & 0.076                      & 0.211                  & 65536                                                     & 1.83E-04        & -        & 22.23            \\
SGS10                           & -                     & 0.010            & 1.13                 & 0.372                      & 0.089                               & 0.101                      & 0.221                  & 131109                                                    & 4.58E-05        & -        & 22.96            \\
SGS11                           & -                     & 0.005            & 1.12                 & 0.396                      & 0.082                               & 0.083                      & 0.036                  & 348741                                                    & 2.29E-05        & -        & 10.51            \\
SGS12                           & -                     & 0.001            & 1.05                 & 0.433                      & 0.128                               & 0.110                      & 0.015                  & 2748011                                                   & 1.43E-06        & -        & 6.10            \\ \cline{1-12}
SGS13                           & 0.041                 & 0.1              & 1.14                 & 0.316                      & 0.044                               & 0.068                      & 0.179                  & 32768                                                     & 3.66E-04        & 3.29     & 17.73            \\
SGS14                           & -                     & 0.05             & 1.15                 & 0.336                      & 0.074                               & 0.089                      & 0.248                  & 53687                                                     & 1.83E-04        & -        & 24.68            \\
SGS15                           & -                     & 0.025            & 1.14                 & 0.342                      & 0.066                               & 0.082                      & 0.185                  & 65536                                                     & 1.83E-04        & -        & 21.00            \\
SGS16                           & -                     & 0.010            & 1.14                 & 0.347                      & 0.070                               & 0.074                      & 0.159                  & 131093                                                    & 4.58E-05        & -        & 19.05            \\
SGS17                           & -                     & 0.005            & 1.11                 & 0.349                      & 0.065                               & 0.089                      & 0.006                  & 345694                                                    & 2.29E-05        & -        & 5.98            \\
SGS18                           & -                     & 0.001            & 1.11                 & 0.395                      & 0.103                               & 0.099                      & 0.010                  & 2752160                                                   & 1.43E-06        & -        & 4.87            \\ \cline{1-12}
SGS19                           & 0.044                 & 0.1              & 1.16                 & 0.297                      & 0.040                               & 0.041                      & 0.187                  & 32768                                                     & 3.66E-04        & 3.50     & 17.17            \\
SGS20                           & -                     & 0.05             & 1.15                 & 0.300                      & 0.039                               & 0.056                      & 0.102                  & 34016                                                     & 1.83E-04        & -        & 12.13            \\
SGS21                           & -                     & 0.025            & 1.15                 & 0.331                      & 0.042                               & 0.051                      & 0.166                  & 65536                                                     & 1.83E-04        & -        & 16.58            \\
SGS22                           & -                     & 0.010            & 1.14                 & 0.335                      & 0.060                               & 0.078                      & 0.152                  & 131093                                                    & 4.58E-05        & -        & 16.16            \\
SGS23                           & -                     & 0.005            & 1.14                 & 0.319                      & 0.069                               & 0.064                      & 0.017                  & 322834                                                    & 2.29E-05        & -        & 8.16            \\
SGS24                           & -                     & 0.001            & 1.14                 & 0.363                      & 0.081                               & 0.086                      & 0.006                  & 2489617                                                   & 1.43E-06        & -        & 5.14            \\ \cline{1-12}
SGS25                           & 0.047                 & 0.1              & 1.17                 & 0.280                      & 0.026                               & 0.045                      & 0.184                  & 32768                                                     & 3.66E-04        & 3.71     & 19.98            \\
SGS26                           & -                     & 0.05             & 1.17                 & 0.291                      & 0.037                               & 0.065                      & 0.211                  & 44219                                                     & 1.83E-04        & -        & 21.34            \\
SGS27                           & -                     & 0.025            & 1.16                 & 0.300                      & 0.044                               & 0.043                      & 0.127                  & 65536                                                     & 1.83E-04        & -        & 13.39            \\
SGS28                           & -                     & 0.010            & 1.13                 & 0.297                      & 0.048                               & 0.064                      & 0.116                  & 131093                                                    & 4.58E-05        & -        & 13.38            \\
SGS29                           & -                     & 0.005            & 1.10                 & 0.298                      & 0.041                               & 0.055                      & 0.005                  & 325608                                                    & 2.29E-05        & -        & 4.48            \\
SGS30                           & -                     & 0.001            & 1.10                 & 0.334                      & 0.067                               & 0.069                      & 0.014                  & 2845525                                                   & 1.43E-06        & -        & 4.49            \\ \cline{1-12}
SGS31                           & 0.05                  & 0.1              & 1.18                 & 0.268                      & 0.022                               & 0.031                      & 0.112                  & 32768                                                     & 3.66E-04        & 3.92     & 12.11            \\
SGS32                           & -                     & 0.05             & 1.18                 & 0.275                      & 0.031                               & 0.050                      & 0.144                  & 32949                                                     & 1.83E-04        & -        & 14.67            \\
SGS33                           & -                     & 0.025            & 1.18                 & 0.289                      & 0.033                               & 0.049                      & 0.164                  & 65536                                                     & 1.83E-04        & -        & 17.65            \\
SGS34                           & -                     & 0.010            & 1.15                 & 0.298                      & 0.040                               & 0.044                      & 0.129                  & 131078                                                    & 4.58E-05        & -        & 16.70            \\
SGS35                           & -                     & 0.005            & 1.14                 & 0.287                      & 0.041                               & 0.044                      & 0.029                  & 315762                                                    & 2.29E-05        & -        & 9.19            \\
SGS36                           & -                     & 0.001            & 1.14                 & 0.312                      & 0.053                               & 0.061                      & 0.005                  & 2481852                                                   & 1.43E-06        & -        & 3.57            \\ \bottomrule
\end{tabularx}
\end{table}
\end{center}

\subsection{Simulations} \label{sec:simulations}

All of our simulations were ran with the $N$-body/SPH code \textsc{gadget-2} \citep{Springel2001,Springel2005}. Other relevant parameters such as number of particles were assigned values that are commonly used in similar works: $N_{\mathrm{d}} = 240,000$ and $N_{\mathrm{h}} = 800,000$, which are the number of particles for disc and halo, respectively. Previous studies indicate that in order to avoid statistical errors in force calculation, the number of particles should exceed $N \approx 10^{6}$ \citep{Dubinskietal2009,Fujii2011}. For generality, we set $c = 10$ and $v_{200} = 160$\,km\,s$^{-1}$ \citep{SpringelWhite1999,Springel2000} which gives a mass of a Milky Way-sized galaxy of $M \simeq 10^{12} M_{\odot}$, with $M_{\mathrm{d}} \simeq 3.5\times10^{10} M_{\odot}$, $M_{\mathrm{h}} \simeq 96.5\times10^{10} M_{\odot}$ for $m_{\mathrm{d}} = 0.035$, and $M_{\mathrm{d}} \simeq 5\times10^{10} M_{\odot}$, $M_{\mathrm{h}} \simeq 9\times105^{10} M_{\odot}$ for $m_{\mathrm{d}} = 0.05$. We also performed additional simulations, tripling $N_{\mathrm{d}}$ and $N_{\mathrm{h}}$ of the original set to verify the results obtained with the low-resolution models (see section~\ref{sec:particle-resolution}).

Table~\ref{tab:t1} summarizes all the simulations conducted here, including the model's parameters and some of the measured quantities. Fig.~\ref{fig:snapshots} shows several snapshots of models that display clear bar formation and Fig.~\ref{fig:snapshots_cont} shows models where bar formation is either transient or non-existent (see section~\ref{sec:results} for details). We discuss their properties later on.

\section{Analysis framework} \label{sec:analysis}

To assess and quantify the models, we carry out kinematic and stability analysis on the stellar discs. We also follow the formation and evolution of bars via the so-called \textit{distortion parameter}, $\eta$ \citep{Shibata2003}, which can be used as a measure of any non-axisymmetric perturbations, and the magnitude of Fourier's second harmonic, $A_{2}$, of the disc surface density specifically captures bar strength.

\subsection{Kinematics} \label{kinematics}
Our kinematic study includes the analysis of the rotational velocity curves and their time evolution for all of our models (see Fig.~\ref{fig:rot_curves}). Velocities in an annulus of the disc are simply the average tangential velocities of all the particles belonging to that annulus,
\begin{equation}
\overline{v}_{\mathrm{t}} = \frac{1}{N_{\mathrm{d}}} \sum_{i=1}^{N_{\mathrm{d}}} v_{\mathrm{t}}^{i},
\label{eq:avg_tan_vel}
\end{equation}
where $v_{\mathrm{t}}^{i}$ is the tangential velocity of particle $i$.

Other quantities, such as the velocity dispersions are used to measure disc heating. Their behaviour gives us clues about the velocity structure once the simulation goes further in time. Fig.~\ref{fig:disp_curves} shows the velocity dispersions for some of the models at different times.  Dispersions are trivially calculated from eq.~\ref{eq:avg_tan_vel}, i.e.
\begin{equation}
\sigma_{i}^{2} = \overline{(v_{i}^{j} - \overline{v}_{i})^{2}} \qquad \mathrm{with} \qquad i = r, \mathrm{t}, z,
\label{eq:dispersions}
\end{equation}
where $v_{i}^{j}$ is the velocity of component $i$ for particle $j$.  $\sigma_{r}$ represents the radial dispersion, $\sigma_{\mathrm{t}}$ is the tangential dispersion and $\sigma_{\mathrm{z}}$ is the vertical dispersion. 
Every plot in Figures~\ref{fig:rot_curves} and~\ref{fig:disp_curves} is measured along the disc's plane, both for simplicity and to correctly assess Toomre's parameter, given that the original conception of this criterion was first applied to infinitely thin discs.

Additionally, in order to quantify the heating we measure the deviations of the velocity dispersions with respect to the initial average values in the radial, tangential and vertical directions:
\begin{equation}
\overline{\sigma}_{\alpha}^{2} = \frac{1}{N_{\mathrm{a}}} \sum_{i = 0}^{N_{\mathrm{a}}} \left( \frac{\sigma_{\alpha,i}^{t} - \sigma_{\alpha,i}^{\mathrm{ini}}}{\sigma_{\alpha,i}^{\mathrm{ini}}} \right)^{2} \qquad \mathrm{with}\,\,\alpha = (r,\mathrm{t},z),
\label{eq:sigma_dev}
\end{equation}
where, for example, $\overline{\sigma}_{r}$ is the deviation from radial dispersion with respect to the IC, and $\sigma_{r,i}^{\mathrm{ini}}$ and $\sigma_{r,i}^{t}$ are the dispersion for every ring in the initial disc and the dispersion for every ring at time $t$ of the simulation, respectively. $N_{\mathrm{a}}$ is the number of rings from a particular snapshot.

\begin{adjustwidth}{-\extralength}{0cm}
\end{adjustwidth}

\begin{figure}
\begin{adjustwidth}{-\extralength}{0cm}
\centering
\includegraphics[scale=0.64]{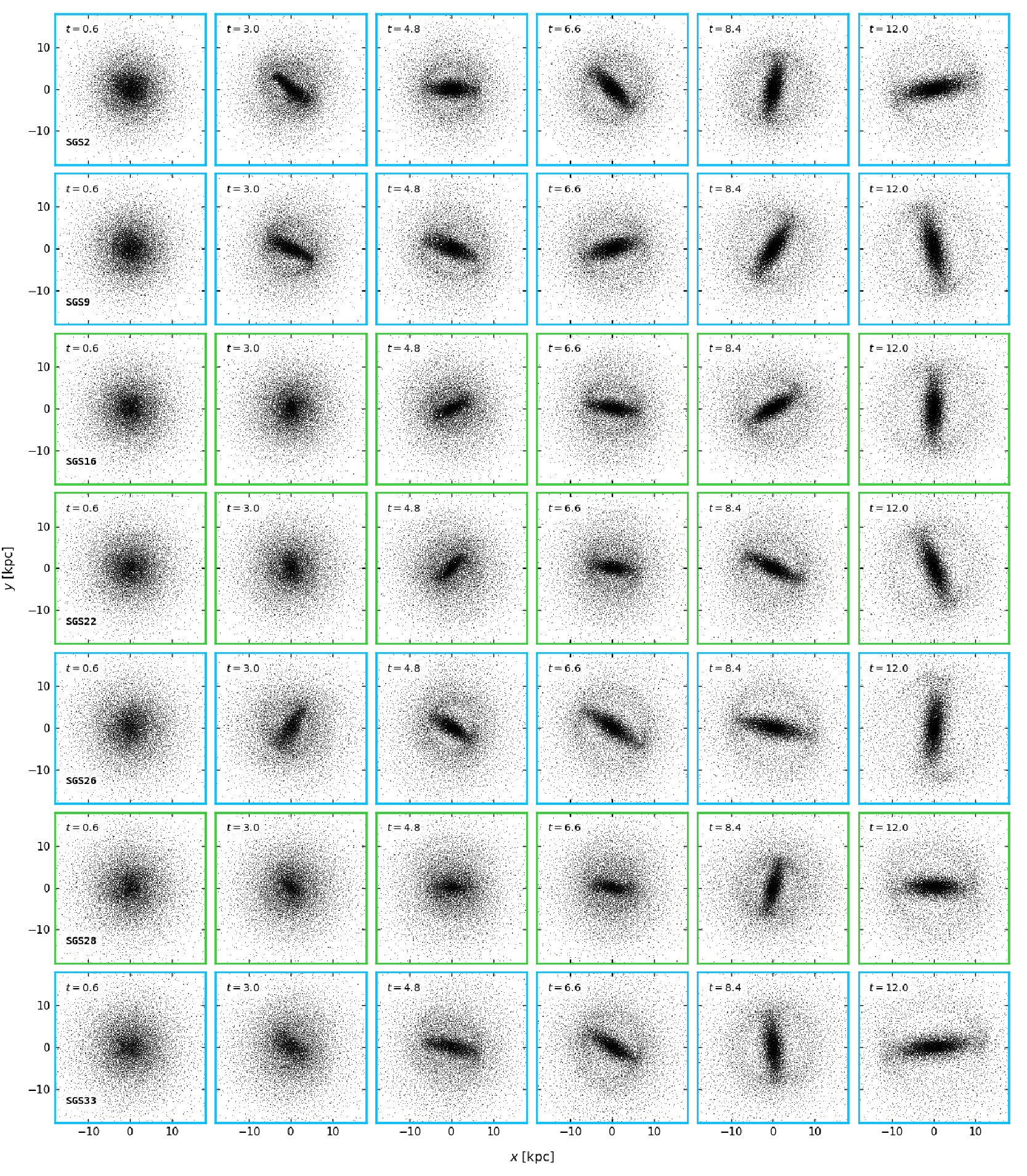}
\end{adjustwidth}
\caption{Disc face-on particle projection for some of simulations from the Table~\ref{tab:t1} that produce strong bars through time. Model name is indicated on the left bottom corner of the first column. Timestamps shown at the top of every snapshot are in units of Gyrs. The frame color of each row indicates: common bar formation \textbf{blue} and delayed bar formation \textbf{green}. See text for details.}
\label{fig:snapshots}
\end{figure}

\begin{figure}
\begin{adjustwidth}{-\extralength}{0cm}
\centering
\includegraphics[scale=0.7]{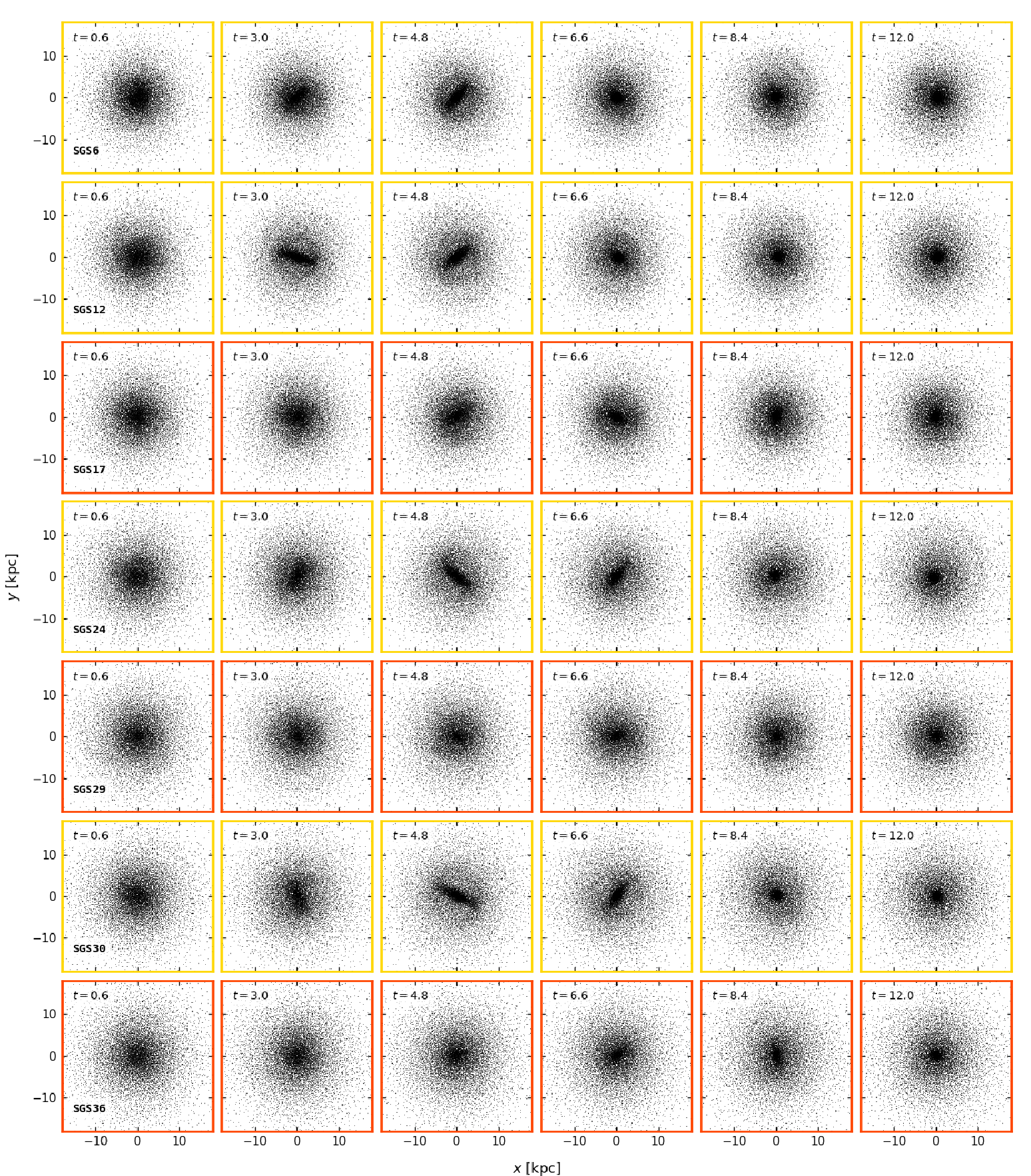}
\end{adjustwidth}
\caption{Same as in Fig.~\ref{fig:snapshots} but this time the simulations show bar stable discs or, in some cases, bar destruction (e.g. SGS12 or SGS30). The frame colors of each row are \textbf{yellow}: transient bar formation and \textbf{red}: no bar formation. See text for details.}
\label{fig:snapshots_cont}
\end{figure}

\begin{figure}
\begin{adjustwidth}{-\extralength}{0cm}
\centering
\sbox0{\includegraphics[scale=0.58]{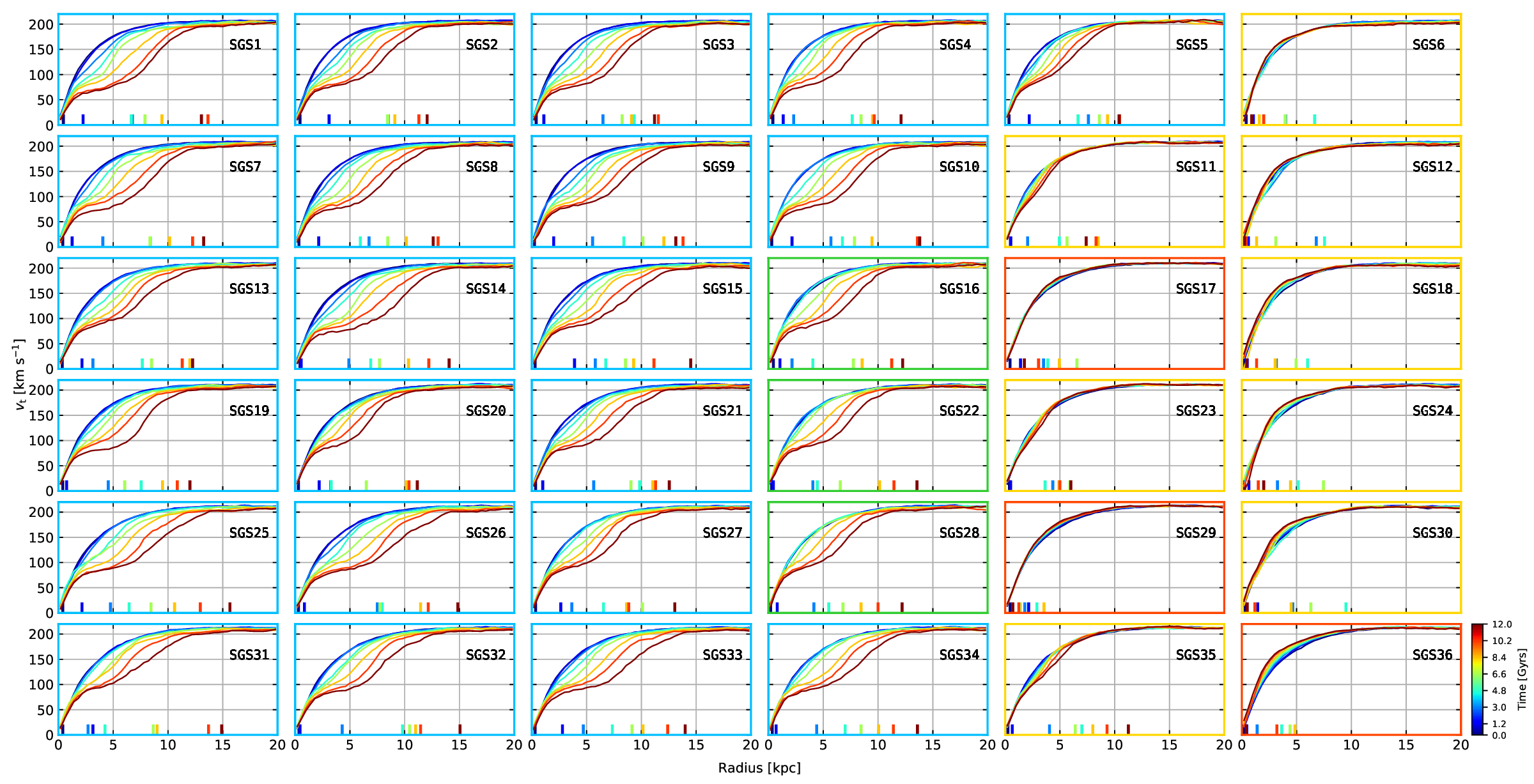}}
\rotatebox{90}{\begin{minipage}[c][\textwidth][c]{\wd0}
\usebox0
\captionof{figure}{Tangential velocities for all models in Table~\ref{tab:t1}. Different colors show the temporal evolution of the tangential velocities indicated by colorbar. The apparent length of the bar is indicated by the colored ticks over the abscissa and the color of each tick matches the color of the corresponding tangential velocity. The color code for each frame represents the same as in Figures~\ref{fig:snapshots} and~\ref{fig:snapshots_cont}. See text for details.}
\label{fig:rot_curves}
\end{minipage}}
\end{adjustwidth}
\end{figure}

\begin{figure}
\begin{adjustwidth}{-\extralength}{0cm}
\centering
\includegraphics[scale=1.27]{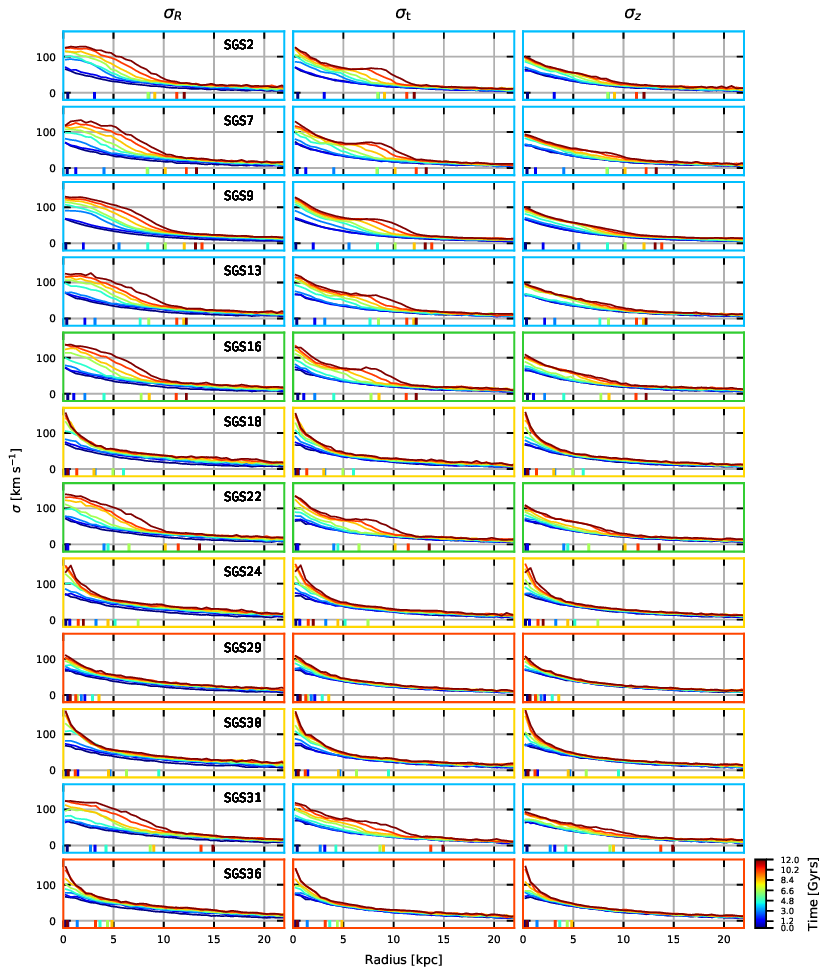}
\end{adjustwidth}
\caption{Velocity dispersion curves for some models from the Table~\ref{tab:t1}. Columns represent --from left to right-- the radial, tangential and vertical dispersions of the models in each row. Different colors show the temporal evolution of each simulation  according to colorbar. The length of the bar is indicated by the colored ticks over the abscissa and the color of each tick matches the color of the corresponding tangential velocity. The color code for the frames represents the same as in Figures~\ref{fig:snapshots} and~\ref{fig:snapshots_cont}. See text for details.}
\label{fig:disp_curves}
\end{figure}	

\subsection{Stability of galactic discs} \label{sec:stability}

\begin{figure}
\begin{adjustwidth}{-\extralength}{0cm}
\centering
\includegraphics[scale=0.63]{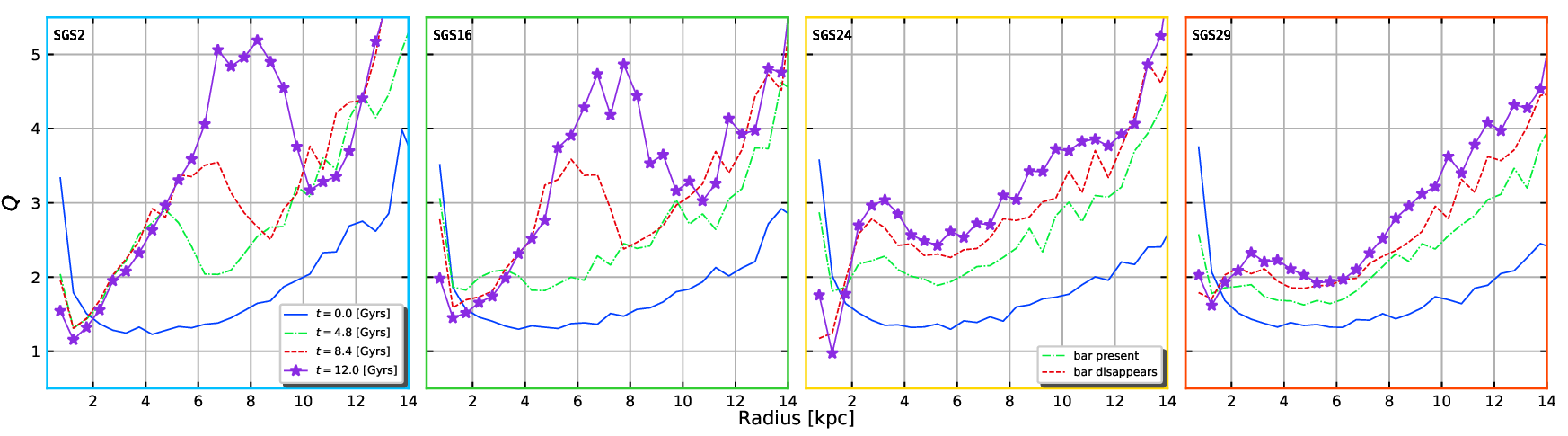}
\end{adjustwidth}
\caption{Toomre's parameter for four different models, each with different degrees of disc distorsion, $\eta_{\mathrm{e}}$. The models with more mass and less softening (SGS24 and SGS29) show the least perturbed discs as can be noted by absence of humps at $R=6-8$ kpc. In the third panel from left to right, we indicate with a legend on the lower right, which line corresponds to a bar being present in SGS24 (green line) and which one corresponds a time where ther is no longer a bar in it (red line). The color code for the frame on each plot represents the same as in Figures~\ref{fig:snapshots} and~\ref{fig:snapshots_cont}.}
\label{fig:Q_evol}
\end{figure}

We take a simple but effective approach to estimate the disc's stability against non-axisymmetric perturbations. We use Toomre's local stability criterion, $Q$ \citep{Toomre1964}, defined by
\begin{equation}
Q = \frac{\sigma_{r}\,\kappa}{3.36\,G\,\Sigma(r)},
\label{eq:toomreq}
\end{equation}
where $\sigma_{r}$ is the radial velocity dispersion, $\Sigma(r)$ is the surface density and $\kappa$ is the epicyclic frequency
\begin{equation}
\kappa = \frac{\partial\,\Omega^{2}}{\partial r} + 4 \Omega
\label{eq:epifrec}.
\end{equation}
Calculating $\sigma_{r}$ and $\kappa$ is relatively easy using equations~\ref{eq:dispersions} and~\ref{eq:epifrec}. Note that this criterion was originally derived for razor thin discs. Because $z_{0} \ll R_{\mathrm{d}}$, we can restrict ourselves to compute $Q$ in the disc's plane \citep{BinneyTremaine2008}. Also, $z_{0} \geq 10\epsilon$ for all of our models, which means that we are able to correctly resolve the vertical structure of the discs. We show the $Q$ profiles and their time evolution for four different types of perturbed discs in Fig.~\ref{fig:Q_evol}.

\subsection{Distortion parameter} \label{sec:distortion-parameter}

Another way to asses the stability of pure stellar discs is by computing the distortion parameter, $\eta$, which was originally used to measure high density concentrations of matter in differentially rotating stars \citep{Shibata2003}. This concept can be applied in a similar manner to a bulk of particles. Obtaining the moments of inertia for the disc we can estimate subtle changes in its density, thus allowing us to know if the distribution exhibits any non-axisymmetric structure. \cite{Gabbasov2006} already applied this method to simulations of isolated galaxy models. They determined that there is a threshold in $\eta$ when simulated discs tend to display signs of bar-like perturbations. In the present work we are not only aiming to replicate and confirm the results of \cite{Gabbasov2006}, but also to expand the usage of $\eta$ to describe properties kindled by the bar, if any.

The distortion parameter $\eta$ is defined as
\begin{equation}
\eta = \sqrt{\eta_{+}^{2} + \eta_{\times}^{2}},
\label{eq:etas} 
\end{equation}
where
\begin{equation}
\eta_{+} = \frac{I_{xx} - I_{yy}}{I_{xx} + I_{yy}} \qquad \mathrm{and} \qquad \eta_{\times} = \frac{2I_{xy}}{I_{xx} + I_{yy}},
\label{eq:np_nt}
\end{equation}
and the moments of inertia are
\begin{equation}
I_{ij} = \sum_{k=1}^{N_{\mathrm{d}}} m_{k} x_{k}^{i} x_{k}^{j} \qquad \mathrm{with} \qquad i,j = (x,y).
\label{eq:moments}
\end{equation}
Here $m_{k}$ is the mass of each particle in \textsc{gadget} units and $x_{k}$ are the positions of every particle in the disc. An important property of $\eta$ is that it allows to track the average movement of every particle. So, in regions where particle density loses symmetry and starts to concentrate in specific parts of the disc, the parameter $\eta$ should increase, thus reflecting the overall motion of the bar component. In section~\ref{sec:results} we demonstrate the different applications of $\eta$. The final value of $\eta$ (called $\eta_{\mathrm{e}}$) for all simulations is shown in Table~\ref{tab:t1}. The threshold $\eta_{\mathrm{e}}$ only shows us whether or not the model holds a bar at the end of the run. Time evolution of $\eta$ will give a full picture of the bar's evolution, as is shown in Fig.~\ref{fig:eta_curves}. In section~\ref{sec:results} we delve into the radial and secular evolution of $\eta$.

\subsection{Fourier magnitude} \label{sec:fourier-mag}

The Fourier magnitude or \textit{strength} is another relevant quantity that describes different stages of bar evolution~\citep[see][]{ValenzuelaKlypin2003}. The magnitudes of different modes for the disc surface density allows to observe how the perturbation grows, whether this growth is rapid or slow, and if other instabilities arise that co-habit with the bar. We measure the strength of the bar as the second harmonic of the Fourier spectra of the disc's particle distribution, that is,

\begin{equation}
A_{2} = \frac{\sqrt{a^{2}_{2} + b^{2}_{2}}}{N_{\mathrm{d}}}.
\label{eq:strength}
\end{equation}

The Fourier components for the second mode are defined as,

\begin{equation}
a_{2} = \sum_{i=1}^{N_{\mathrm{d}}} \cos(2\theta_{i}) \qquad b_{2} = \sum_{i=1}^{N_{\mathrm{d}}} \sin(2\theta_{i}).
\label{eq:components}
\end{equation}

where $\theta_{i}$ is the polar angle of particle $i$. In order to obtain representative values of the strength, we take the maximum of $A_{2}$ at each time, which is a valid description of bar amplitude~\citep{SahaNaab2013}. Evolution of bar strength for some models is given in Fig.~\ref{fig:strength}.

\section{Results and Discussion} \label{sec:results}

We now describe and analyse the results obtained throughout this work. Table~\ref{tab:t1} summarizes some of the quantities extracted from all the simulations. 

From these data we notice that $Q_{\mathrm{min}}$ (column 4) is generally higher as $m_{\mathrm{d}}$ increases. This is true when comparing models with the same $\epsilon$ (e.g. SGS1 and SGS31 ). We also note that is unclear how disc stability depends on $\epsilon$. When $\epsilon \leq 0.005$ and $m_{\mathrm{d}}$ remains the same, $Q_{\mathrm{min}}$ reaches its lowest values. However, for the rest of models (which have $\epsilon \geq 0.01$) $Q_{\mathrm{min}}$ is practically the same. 

Now, $Q_{\mathrm{min}}$ tells us that, generally, simulations with greater disc mass tend to be more stable, which is not a surprising result \citep{AthanassoulaMisiriotis2002,Athanassoula2013}, given that disc dominated models are less prone to manifest structures such as bars. In our models, the increase in $Q_{\mathrm{min}}$ is mostly due to an increase in the radial dispersion for higher values of $m_{\mathrm{d}}$. The models also increase in $R_{\mathrm{d}}$, but this quantity does not affect the behavior of $Q_{\mathrm{min}}$ because both surface density and epicyclic frequency, and their effects, more or less cancel out when $R_{\mathrm{d}}$ increases. Such an increase in the radial dispersions may be explained by enhanced kinematic pressure in the disc due to the fact that increasing $m_{\mathrm{d}}$ actually reduces the central surface density of discs, which lowers the value of $Q_{\mathrm{min}}$. The latter requires a thorough analysis that is out of the scope of this work. The dependence on $\epsilon$ is much more noticeable: for smaller values of $\epsilon$, $Q_{\mathrm{min}}$ decreases, making the disc theoretically more responsive to local instabilities. However, we must note that the dependency between $m_{\mathrm{d}}$, $\epsilon$ and $Q_{\mathrm{min}}$ is small since basically all of the models have $Q_{\mathrm{min}} \approx 1.1$. Thus, we may regard the models as being marginally stable against bar formation.
 
The four models in Fig.~\ref{fig:Q_evol}, SGS2, SGS16, SGS24 and SGS29 have different masses and softenings. It is evident that SGS2 and SGS16 are very similar, structurally speaking and, as shown in Figures~\ref{fig:snapshots} and~\ref{fig:rot_curves}, they effectively exhibit a bar. Model SGS24 seems to have a bump at $t = 4.8$\,Gyrs in $Q$ and, if we look carefully in Fig.~\ref{fig:snapshots_cont}, it could host a bar in an early stage around the same time, but is nowhere to be found in later snapshots (see also model SGS30 in Fig.~\ref{fig:snapshots_cont}). Model SGS29 exhibits a similar bump at $t = 8.4$\,Gyrs, but there is no bar in either its particle snapshots or velocity curves. This bump in SGS29 is caused by the gathering of particles at its centre, something that also occurs in models that have transient bars, like SGS24. We offer an explanation for this phenomenon in section~\ref{sec:vertical-acc-profiles}. 

Columns 5, 6 and 7 in Table~\ref{tab:t1} are the deviations from average dispersion velocities, $\bm{\overline{\sigma}_{r}}$, $\bm{\overline{\sigma}_{\mathrm{t}}}$ and $\bm{\overline{\sigma}_{z}}$ in the radial, tangential and vertical directions, respectively.
The deviations are evaluated at $t = 1.8$\,Gyrs, time where the first bar just started to form\footnote{We use this time as a reference point to assess the involvement that bar ignition has on any increase in the velocity dispersions (``heating'').}(model SGS5). The values given in the table are interpreted as follows. If, for instance, $\overline{\sigma}_{\alpha} = 1$, it means that, for coordinate $\alpha$, the average dispersion velocity at the end of the simulation is $2$ times higher compared to that of the IC. Conversely, if $\overline{\sigma}_{\alpha} = 0$, the average dispersion has remained the same throughout the simulation. In this manner, we can measure the disc heating due to selected attributes rather than caused by the bar instability.

Column 8 is the distortion parameter $\eta_{\mathrm{e}}$ and captures the disc symmetry at the end of the simulation and columns 9 and 10 give the average timestep and the total number of steps per simulation. The tendency for the timestep to decrease when $\epsilon$ decreases is expected due to the presence of particles with higher accelerations. 

\begin{figure}
\begin{adjustwidth}{-\extralength}{0cm}
\centering
\includegraphics[scale=0.43]{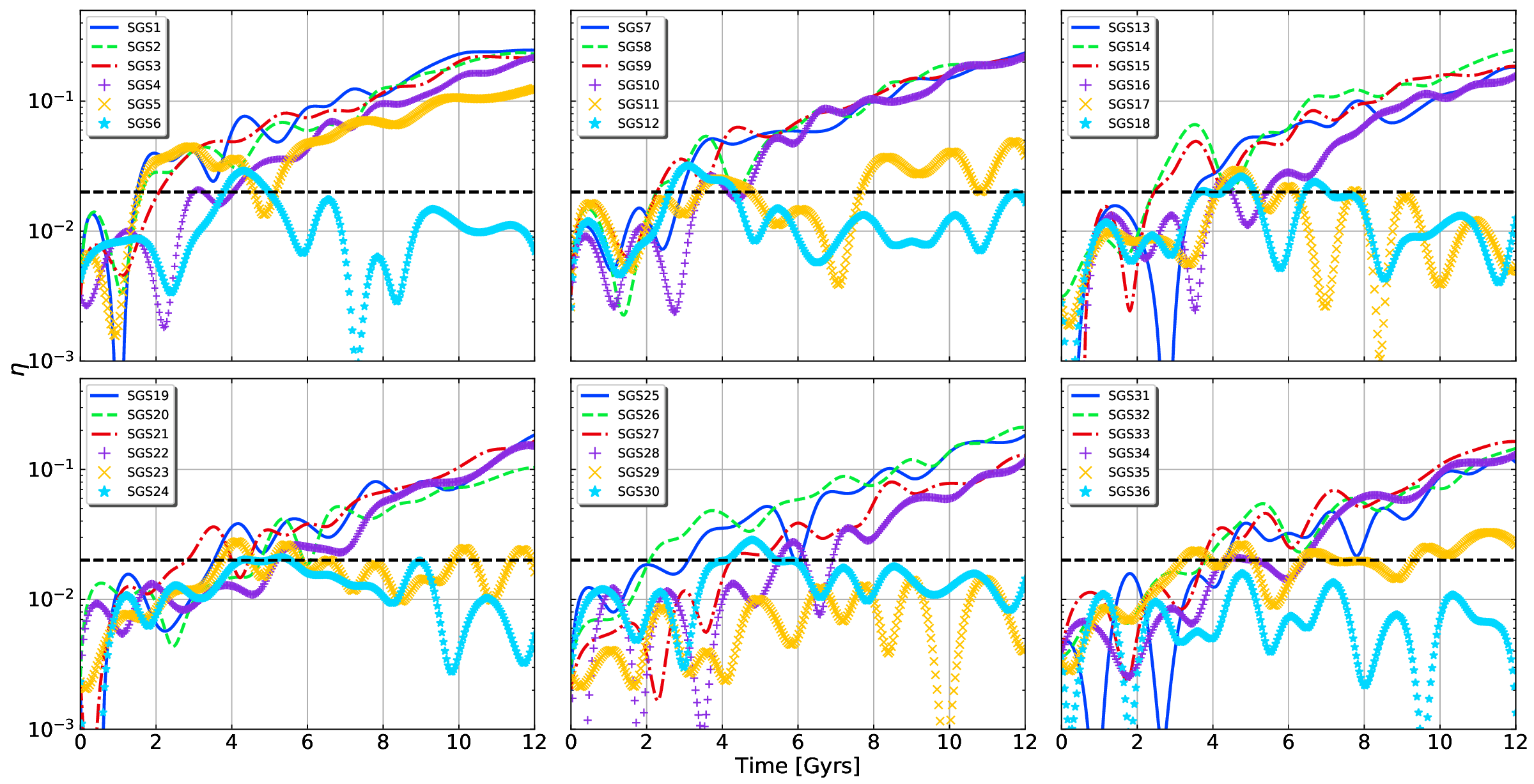}
\end{adjustwidth}
\caption{Time evolution of $\eta$. The dashed black line represents $\eta_{\mathrm{th}} = 0.02$, which is the threshold for definite bar formation.}
\label{fig:eta_curves}
\end{figure}

Figures~\ref{fig:snapshots} and~\ref{fig:snapshots_cont} exemplify disc evolution for models that exhibit barred discs and transient/non-barred discs, respectively. Only $10{\%}$ of disc particles are shown in order to correctly distinguish the bars. We identify, from both Figures~\ref{fig:snapshots} and~\ref{fig:snapshots_cont}, and the distorsion parameter in Figure~\ref{fig:eta_curves}, four scenarios of bar formation and color-code plot frames according to the following:
(i) \textbf{blue}: common bar formation, where the bar is clearly formed and there is no comparative delay in its growth;
(ii) \textbf{green}: delayed bar formation, at least in comparison to the common scenario;
(iii) \textbf{yellow}: transient bar formation, where at some point the bar forms but then it is destroyed;
(iv) \textbf{red}: no bar formation, for cases where no clear distortion was detected at any time.
Models in Fig.~\ref{fig:snapshots} display strong persistent bars, where it can be noted that the bar triggering varies in time. On the contrary, models in Fig.~\ref{fig:snapshots_cont} show either stable discs or weak bars and their destruction. Visual comparison of bar growth may already give some clues on the influence of the model parameters. For example, model SGS26 produces the bar more rapidly than model SGS28 and, given that both models have the same properties except $\epsilon$, we might conclude that the amount of softening could have affected the bar growth rate. Models with the lowest $\epsilon$ show little signs of perturbation, at least visually. The cases of SGS24 and SGS30 especially stand out because there is a barred structure forming at $t \geq 3.0$\,Gyrs, but it then disappears at $t \geq 7.2$\,Gyrs. These two models only differ in disc mass by ${\sim}7$\%. We further confirm these results through the measurement of distortion parameter in section~\ref{sec:evol_eta}.

Snapshots of models SGS30 and, in a fainter fashion, SGS24, in Fig. \ref{fig:snapshots_cont} show traces of bar creation and destruction (or perhaps supression) in a period of about 3~Gyrs. The self-consistent vanishment of bars in these models is not due to matter concentration or the accumulation of eccentric orbits~\citep[e.g.][]{Normanetal1996,Guedesetal2013,Kormendy2013} because our models do not have gas to gather and the dispersion curves do not reveal any strange behaviour, such as extreme vertical heating. It is in fact possible that we are dealing with a numerical artifact. In such case, we could use statistical tools designed to unmask spurious effects \citep[e.g.][]{Athanassoulaetal2000,Gabbasov2006}. The most useful is perhaps the estimation of errors in force calculation~\citep{Dehnen2001}. Unfortunately, the complex nature of disc-halo dynamics imposes serious challenges for them to be applied appropriately. This is because quantities such as the Average Square Error (ASE) and Integrated Square Error (ISE) are commonly calculated for density distributions that are spherically symmetric \citep{Merritt1996,Athanassoulaetal2000}, which is not the case for our simulations since the density of our DM halos has been affected by the presence of the bar. Nonetheless, we performed several realizations of a model with an isolated DM halo (no disc) to compute ASE in order to determine their optimal softening length. The base model has $N_{\mathrm{h}} = 800,000$, a total mass of $M_{200} \approx 10^{12} M_{\odot}$, scale length $a = 27.06$ kpc and $c=10$. We created 24 realizations of this model with a softening length window of $0.001$ $\leq \epsilon \leq 3$. The forces associated with the halo realizations are calculated using the cubic spline and are then compared to the actual forces produced by the Hernquist profile through the ASE measurement. The value of the resulting errors is normalized with a factor defined by $C = 1/F_{\mathrm{max}}^2$, where $F_{\mathrm{max}}$ is the maximum force exerted by the theoretical halo density distribution~\citep{PriceMonaghan2007,DasDebBaruah2021}. We determined an optimal Plummer softening length of $\epsilon_{\mathrm{opt}} \approx 0.3$, which falls outside of our original softening window. However, we are interested in the behavior of the disc subjected to changes in particle and space resolution, and the disc density is higher than that of the halo. Furthermore, the halo in our models is affected by adiabatic contraction due to its interaction with the disc, so we expect it to have a denser core than in the isolated case. All of this means that the optimal softening for the models should be considerably smaller than the one found for the realizations.

We estimated the relaxation time via two methods described in~\cite{Athanassoula2001}: through
	
\begin{equation}
T_{\mathrm{relax}} = \frac{N}{8\ln(R/\epsilon)} T_{\mathrm{cross}},
\label{eq:relaxation-time1}
\end{equation}

where $R$ is the whole halo extension and $T_{\mathrm{cross}}$ is the crossing time of the system, and through the derived fitting equation that depends on the number of particles

\begin{equation}
\log_{10}(T_{\mathrm{relax}}) = 0.63 + 0.78 \log_{10}(N)
\label{eq:relaxation-time2}
\end{equation}

Using equation~\ref{eq:relaxation-time1}, we find that all of our models have a relaxation time that exceeds ${\sim}10^{12}$ years. However, if we calculate the relaxation times using equation~\ref{eq:relaxation-time2} assuming a model with only a halo component (with a $\gamma=0$ Dehnen profile and $\epsilon=0.27$ kpc), we find that, for $N=8\times 10^5$ particles, $T_{\mathrm{relax}}\sim 10^5$ years; for $N=2.4 \times 10^6$ particles we also find relaxation times of ${\sim}10^5$ years. So, equation~\ref{eq:relaxation-time1} seems to overestimate the relaxation times and equation~\ref{eq:relaxation-time2} seems to underestimate them. We are not entirely sure about the applicability of the fitting equation to a broader family of models (e.g. Hernquist or NFW profiles), but there are several examples of numerical studies of galaxy models that already use softening lengths in the range we are examining, with similar number of particles and mass resolutions~\citep{Dubinskietal2009,SahaNaab2013,Collier2020}, that appear to be physically realistic.

\subsection{Bar evolution on velocity curves} \label{sec:bar-evol-velocity-curves}

The model depicted in Fig.~\ref{fig:rot_curve_init} and described in section~\ref{sec:ics} has a velocity structure that mirrors a rigid body in its central regions (high dispersions) and a differentially rotating body in its outer regions (low dispersions). 
After just a few rotation periods, the models in Fig.~\ref{fig:rot_curves} already show effects induced by a bar-like perturbation. The shapes of our barred tangential velocities do resemble some of the observed tangential velocities for barred galaxies. For example, \cite{FathiBeckmanetal2009} measured, among other things, the velocity curves of 10 late-type barred galaxies. The tangential velocities of their stronger bars --for instance, NGC 7741 or NGC 7479-- are quite similar to what we obtained in our simulations. When a bar appears, the velocity profile in such region should be similar to a rigid body; as the bar grows this effect strengthens. In our simulations the bar seems to occupy almost all of the disc, which is shown by the coloured bar length ticks in Fig.~\ref{fig:rot_curves}. The lengths are determined by following the phase of the bar, which can be found using the Fourier magnitudes described in section~\ref{sec:fourier-mag}, i.e. $\phi_{\mathrm{b}} = \arctan(b_{2}/a_{2})$. The bar is found in regions where $\phi_{\mathrm{b}}$ remains constant; the length of the bar is defined as the annulus where $\phi_{\mathrm{b}}$ deviates $\pm \arcsin(0.3)$ from its constant value (see \cite{AthanassoulaMisiriotis2002} for a detailed description of this method). The drawback of this method is that it is reliable only for strong bars where their phase can be uniquely determined.

From the rotational velocity curves and velocity dispersions in Figures~\ref{fig:rot_curves} and~\ref{fig:disp_curves}, respectively, we can notice some characteristic features ignited by a bar-like perturbation. For all the models showing a bar, the tangential velocities exhibit a decrease in velocity for central regions of the disc. This is because bars considerably change the density distribution of discs, in such extent that they are able to produce eccentric orbits (commonly known as the $x_{2}$ family orbit) in the centre of the disc, which are the main constituents of bulge components \citep{Normanetal1996,Kormendy2013}.
Knee-shaped curves at later stages of evolution indicate the existence of four different regions: a central solid body region, a differentially rotating plateau, external solid-body region, and external differentially rotating disc. It is important to note that the shape of the tangential velocities in the central and external regions barely change throughout the simulation.

Comparing the velocity dispersion curves, it is evident that radial and tangential dispersions reflect the same correlation with the bar strength as seen in the tangential velocities, forming a bump in the centre for models with a bar present. At the same time, vertical dispersions --commonly used to measure the degree of the disc heating-- show little change when subjected to the bar. For comparison, the models with vanishing or no apparent bar (SGS24, SGS29, SGS30, and SGS36) show an increase only in the centre of the disc. We expect that models with small softening values to experience disc heating (in all directions), since these models tend to become collisional. We see no clear evidence of such effect, other than the central region. We may conclude that the main factor responsible for the disc heating in our models is the bar instability.

Looking at Fig.~\ref{fig:disp_curves}, we can notice a pattern for $\sigma_{z}$: as $\epsilon$ decreases, discs tend to get hotter only in their inner-most region (models SGS24, SGS30, and SGS36). SGS24 shows no trace of a bar in the finishing stages of its evolution, but Figures~\ref{fig:rot_curves}, \ref{fig:disp_curves} and~\ref{fig:eta_curves} show, in earlier stages ($t \simeq 4$ Gyrs), signs of bar formation. This means that the perturbation was suppressed spontaneously without the need of an external instigator. There is also a noticeable discrepancy between the dispersion curves of SGS29 and SGS36. Both models never formed a bar; however, SGS36 displays considerable amount of central heating in all velocity components while SGS29 does not; this effect could be related to changes in either $m_{\mathrm{d}}$ or $\epsilon$.

\subsection{Evolution of parameter $\eta$} \label{sec:evol_eta}

\begin{figure}[t]
\centering
\includegraphics[scale=0.85]{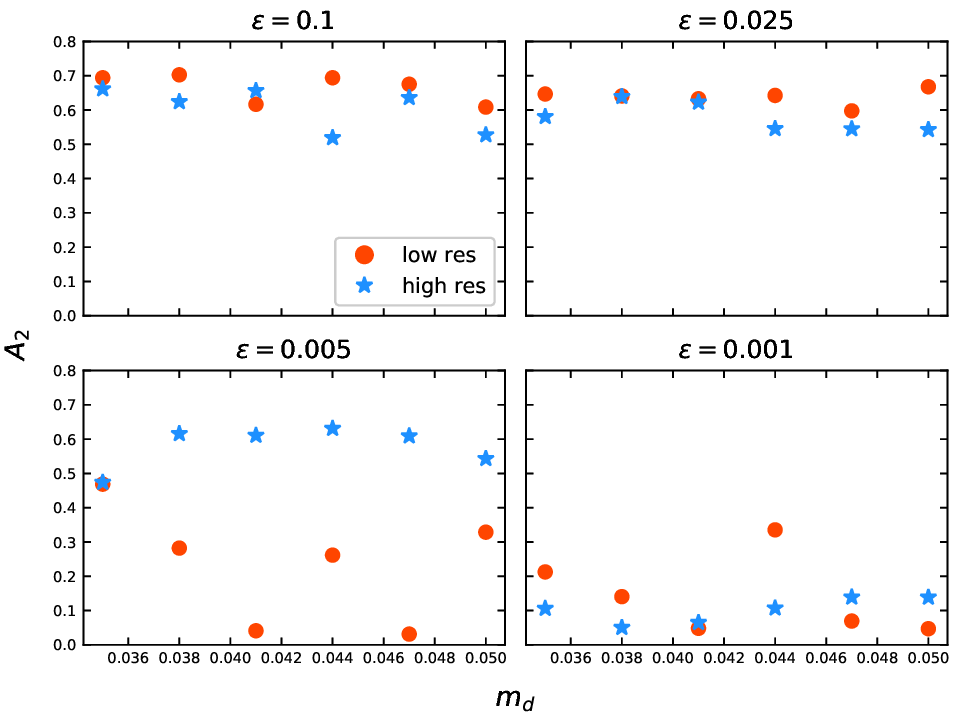}
\caption{$\max(A_{2})$ dependence on $m_{\mathrm{d}}$ for constant values of $\epsilon$ at $t = 12$ Gyrs. The symbols represent the same as in Fig.~\ref{fig:relations_mass}.}
\label{fig:relations_soft}
\end{figure}

\begin{figure}
\centering
\includegraphics[scale=0.85]{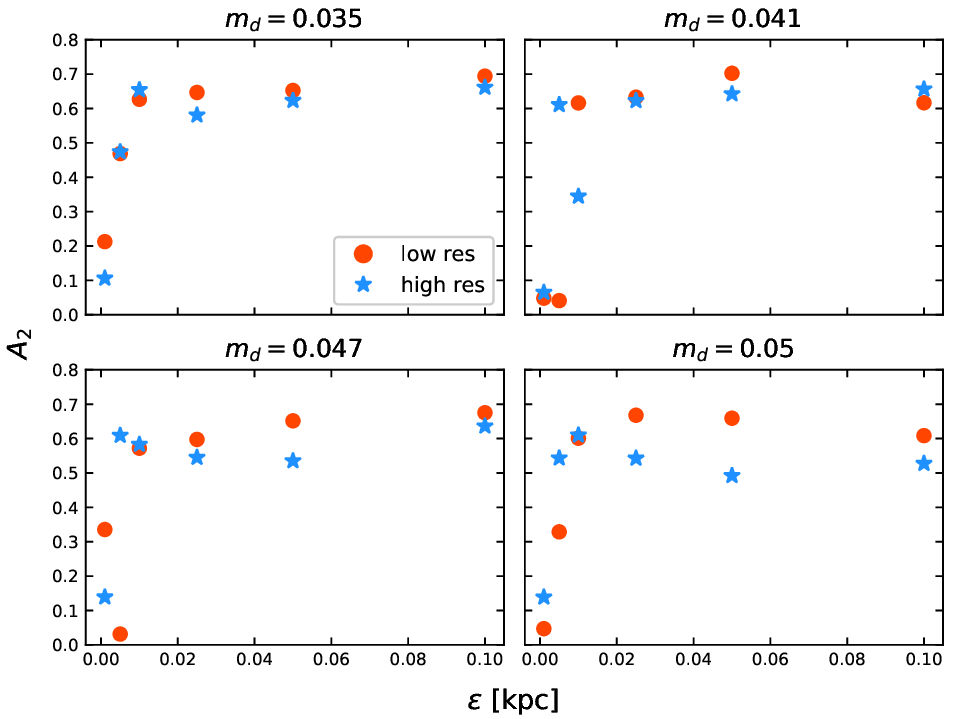}
\caption{$\max(A_{2})$ dependence on $\epsilon$ for constant values of $m_{\mathrm{d}}$ at $t = 12$ Gyrs. The models marked with a red bullet ({\Large$\bullet$}) are of high resolution, while those with a filled blue star ($\filledstar$) are of low resolution.}
\label{fig:relations_mass}
\end{figure}

Fig.~\ref{fig:eta_curves} shows $\eta$ values evolving through time for all of our models, giving a more complete view of disc and bar behaviour. It is important to mention that all of the $\eta$ curves in Fig.~\ref{fig:eta_curves} were smoothed with a cubic kernel interpolation, simply to reduce the excess of noise in the curve caused by oscillations in the position of the disc's centre of mass.

According to \cite{Gabbasov2006}, when $\eta \geq \eta_{\mathrm{th}} = 0.02$, the disc is set to form  perturbations similar to bars. Our experiments confirm this threshold, even for transient bars. Most of the models do have clear bar presence (compare Figures~\ref{fig:snapshots} and \ref{fig:eta_curves}) and all of these models surpass the prescribed threshold of $\eta_{\mathrm{th}} \simeq 0.02$. Moreover, all the models with undisputed bar formation have $\epsilon \geq 0.01$. On the other hand, the effect of the disc mass is indeed small due to the narrow range used for this parameter, although it is still noticeable. When $m_{\mathrm{d}}$ increases, the time the simulation takes to form its bar is slightly longer (compare upper-left and lower-right panes in Fig.~\ref{fig:eta_curves}), from ${\sim}2$ to ${\sim}4$ Grys. Once a simulation exceeds the value of $\eta_{\mathrm{th}}$,  the bar recently created keeps on gaining momentum, accumulating more mass (particles), and thus increasing its overall strength. Results for $\epsilon \leq 0.005$ are puzzling. For the lowest values of softening, the bar instability suffers to keep itself alive; in fact, the only model with $\epsilon \leq 0.005$ that maintains its bar growing is SGS5, which also has the smallest possible value of disc mass in our set ($m_{\mathrm{d}} = 0.035$). Models SGS29 and SGS36 are the only ones that never exhibit signs of clear distortion. Model SGS17 surpasses $\eta_{\mathrm{th}}$ at $t \sim 4$\,Gyrs, but there is no clear visual confirmation so we classified it as an unbarred model. The rest of the models with these softening values (SGS6, SGS11, SGS12, SGS18, SGS23, SGS24, SGS30, and SGS35) show transient or unclear bar formation. For example, SGS24 gathers enough particles to elevate $\eta$ beyond the threshold of 0.02 in ${\sim}4$ Gyrs and then, ${\sim}1$ Gyr later, loses such mass and is never capable to recover its (faint) bar. The mechanism that describes such bar supression ought to be different from the ones propose by \cite{Normanetal1996,Berentzenetal2004,SahaElmegreen2018}. An earlier explanation for this effect was first exposed by \cite{Sellwood1981}, pointing that $\epsilon$ could play a paramount role in the suppression of bar-like instabilities. Later on, \cite{Romeo1994} concluded that, for simulations of one-component galaxy i.e., stars only, and extremely low values of softening lengths the bar instability was considerably suppressed. The degree of suppression depended on how small the softening was. \cite{Romeo1994} also find that bar suppression correlates to the size of the perturbation, $\lambda$; if $\lambda < \epsilon$ then the disc will hardly react to it.

Fig.~\ref{fig:eta_curves} confirms that the model showing the least perturbed disc also has the most softened potential (SGS36). Unfortunately, this conclusion is not definitive, since models with the lowest $\epsilon$ are not always the least perturbed (see $\eta$ curve in Fig.~\ref{fig:eta_curves} for models SGS29 and SGS30, respectively). 

These results may lead to the assumption that if one decides to increase $m_{\mathrm{d}}$ or decrease $\epsilon$ independently, the simulated discs would follow a trend: increasing $m_{\mathrm{d}}$ --or decreasing $\epsilon$-- heats up the disc and increases its stability; decreasing $m_{\mathrm{d}}$ or increasing $\epsilon$ does the opposite. In this case, distinctive pairs of values for $m_{\mathrm{d}}$ and $\epsilon$ seem to resonate with each other, i.e. either strengthen or diminish the bar. Furthermore, models SGS29 and SGS35 both have the same softening length but SGS35 has a bit more disc mass, so we would expect to witness less distortion in SGS35, which is not true (see their respective $\eta$ curves). Another example is model SGS26. Comparing this model with others of the same softening but less disc mass (e.g. SGS20), we find that SGS26 starts forming its bar earlier. A clear-cut explanation for this non-linear behaviour is tough to find. It makes sense that $m_{\mathrm{d}}$ and $\epsilon$ interact with each other: an increase of disc mass also increases the mass of individual particles, which alters force computation. In consequence, we must take into account these changes in the force field in order to choose an appropriate value for $\epsilon$. 

\begin{figure}
\centering
\includegraphics[scale=0.71]{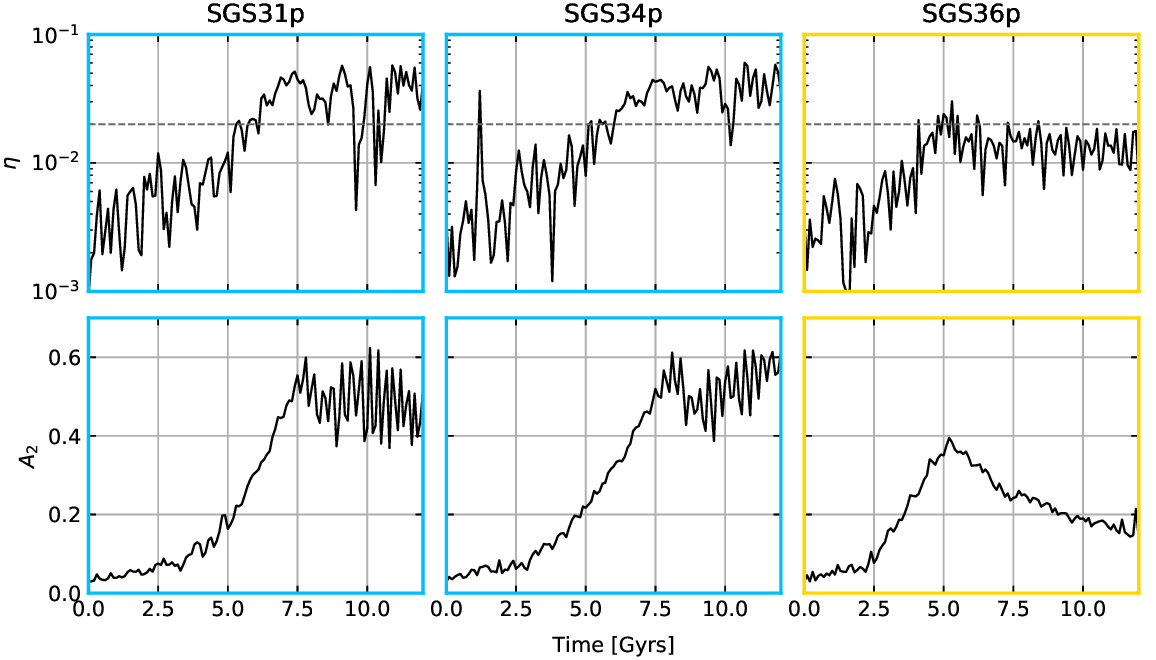}
\caption{Time evolution of $\eta$ (top) and the strength $A_{2}$ (bottom) for the discs of models SGS31p (left), SGS34p (centre) and SGS36p (right). The horizontal dashed line in the $\eta$ plot is our bar threshold, $\eta_{\mathrm{th}} = 0.02$. The color code for the frame in each plot represents the same as in Figures~\ref{fig:snapshots} and~\ref{fig:snapshots_cont}.}
\label{fig:strength}
\end{figure}

\subsection{Particle resolution} \label{sec:particle-resolution}

An important factor on the reliability of $N$-body simulations is the particle resolution. We have already established in section~\ref{sec:introduction} that $\epsilon$ and $N$ are intimately related~\citep[e.g.][]{Merritt1996,Dehnen2001}, no matter the scale~\citep{IannuzziDolag2011,Zhangetal2019}. Improper balance between these parameters may result in undesired outcomes. For a given $N$, small softening values might cause undesired relaxation (heating) which could damp the generation of local instabilities, either barred or spiral~\citep{Zhangetal2019}. Conversely, high softening values may introduce large biases that could render simulations unrealistic.

This means that, for the range of softening lengths used here, it may be expected that some of the results presented above show erratic behaviour. Following our previous analysis, we will consider the next relationships between $\eta$ and the numerical parameters $m_{\mathrm{d}}$ and $\epsilon$: $\eta \propto m_{\mathrm{d}}^{-1}$ and $\eta \propto \epsilon$, and the models that conform to the above are labelled as ``well-behaved'' or normal. These relations are not followed by all of our models. For instance, models SGS14 and SGS26 have longer and stronger bars than models with larger softening values, e.g. SGS13 and SGS25 (see Figures~\ref{fig:rot_curves} and~\ref{fig:eta_curves}), despite all having the same amount of mass. On the contrary, models SGS8 and SGS20 with $\epsilon=0.05$ have shorter and weaker bars than SGS9 and SGS21 with $\epsilon=0.025$. These results may imply a connection between parameters $\epsilon$ and $m_{\mathrm{d}}$, given that they do not present normal behaviour.

To verify this is not a coincidence or randomness within our simulations due to the low number of particles or range of softenings, we have ran some of our models again, this time tripling the number of particles for each component, which means that $N_{\mathrm{d}} = 0.72\times 10^{6}$ and $N_{\mathrm{h}} = 2.4\times 10^{6}$, leaving all the other quantities untouched. We added the label {\bf `p'} to these models to distinguish them from our original set. The resulting properties of the discs in these experiments are summarized in Table~\ref{tab:t2}. 
Comparing $Q_{\mathrm{min}}$ of this table and the one in Table~\ref{tab:t1}, we immediately notice that the models with more particles have hotter discs than the fiducial ones, and hence they are more stable. This fact is reflected by $\eta_{\mathrm{e}}$, which is lower for almost all of the models in Table~\ref{tab:t2}, compared to those in Table~\ref{tab:t1}. In our low resolution models, the strength of bars is not necessarily connected to Toomre's stability criterion, which is self-evident when comparing the values of $Q_{\mathrm{min}}$ and $\eta_{\mathrm{e}}$ for models with $\epsilon \leq 0.005$, as discussed in the beginning of section~\ref{sec:results}. 
Furthermore, $\epsilon$ and $m_{\mathrm{d}}$ still show a non-linear influence in regards to bar formation and growth, e.g. SGS2p reaches the same bar strength ($\eta_{\mathrm{e}} \approx 0.12$) as SGS29p, despite having less disc mass and a lower softening length (see Table~\ref{tab:t2}). 

As mentioned earlier, $\epsilon$ and $N$ are closely related. This was already studied for some simple configurations with analytically known density-potential pairs~\citep[e.g][]{Merritt1996,Athanassoulaetal2000}. So, for a Plummer sphere and a fixed value for $\epsilon$, the total error decreases with $N$, independently from the choice of $\epsilon$~\citep{RodionovSotnikova2005,Zhangetal2019}. Consequently, our extra set of simulations should be physically more faithful than our fiducial one, regardless of our parameter choices. By increasing $N$, we are essentially moving the softening length window so that force calculation for particles found at short distances (e.g $\epsilon \approx 0.005$) is closer to the `real' interacting force, instead of using a lower force magnitude to avoid divergence in the accelerations~\citep{Springel2001}. A similar effect may be achieved by setting the softening length of halo particles so that the force between these and disc particles is the same. In appendix~\ref{sec:apen1}, we study the latter scenario on model SGS5 more closely.

Further evidence that there is an interplay between $\epsilon$ and $m_{\mathrm{d}}$ can be found in Figures~\ref{fig:relations_soft} and~\ref{fig:relations_mass}. For each panel in Figure~\ref{fig:relations_soft}, a fixed value of softening is chosen and $\max(A_2)$ is plotted against disc mass fraction for both the low-resolution (LR) and high-resolution (HR) models. For the largest softening value considered in this work ($\epsilon = 0.1$; top left panel), the value of the bar strength remains approximately constant, $\max(A_2) \sim 0.675 \pm 0.025$, with the exception being $m_{\mathrm{d}} = 0.044$. For this $m_\mathrm{d}$ value, the LR model has $\max(A_2) \sim 0.7$ while for the HR model, $\max(A_2) \sim 0.5$. For a softening value of 0.025 (top right panel), the bar strength remains approximately constant, $\max(A_2) \sim 0.6 \pm 0.05$, regardless of disc mass fraction or numerical resolution. For a softening value of 0.005 (bottom left panel), the LR and HR models show important discrepancies in the bar strength (larger than 0.43 between HR and LR models) except for the lowest value of disc mass fraction. This might be pointing out a limiting $(\epsilon, m_{\mathrm{d}})$ value for which the models can be considered as physically truthful using a low resolution, $N \sim 10^6$; meaning that, for $m_{\mathrm{d}} \geq 0.038$, LR models with $\epsilon = 0.005$ cannot be trusted. For the lowest softening value ($\epsilon = 0.001$), $\max(A_2)$ values show no tendency with disc mass fraction regardless of
the resolution, and may be an indication that a softening value of $\epsilon = 0.001$ holds no physical validity, regardless of the $m_{\mathrm{d}}$ or $N$ values.

On the other hand, for each panel in Figure~\ref{fig:relations_mass}, $\max(A_2)$ is plotted against softening for a fixed value of
$m_\mathrm{d}$. When $\epsilon \geq 0.025$, all models form a bar with a strength of $\max(A_2) \sim 0.6 \pm 0.1$. For $\epsilon = 0.01$, the strengths only differ when $m_{\mathrm{d}} = 0.041$. For $\epsilon = 0.005$, the differences become much more significant, having LR and HR pair models with a clear bar ($m_{\mathrm{d}} = 0.035$; $\max(A_2) \sim 0.5$), another pair with a strong bar in the HR model ($m_{\mathrm{d}} = 0.047$; $\max(A_2) \sim 0.6$) and no bar at all in the LR model ($m_{\mathrm{d}} = 0.047$; $\max(A_2) \sim 0$). In fact, all HR models with $\epsilon = 0.005$ develop a bar. This results suggest a threshold in terms of $\epsilon$ for bar formation in LR models. Such threshold is also affected by the value of $m_{\mathrm{d}}$. When $\epsilon = 0.001$, none of our models develop a bar, regardless of the resolution. Considering that LR models appear to have a bar threshold, it is also fair to suggest that the same happens for HR models, but this threshold is moved by increasing $N$. We resume this analysis in section~\ref{sec:vertical-acc-profiles}.

Fig.~\ref{fig:strength} shows the temporal evolution of $\eta$ (top row) and the $A_{2}$ Fourier component (bottom row) for models SGS31p, SGS34p and SGS36p. We chose these models because they are the most stable in terms of their $Q_{\mathrm{min}}$. Also, their stability criterions are basically the same, so any differences among them would be related to $m_{\mathrm{d}}$ and/or $\epsilon$. Both SGS31p and SGS34p have, in essence, the same behaviour: they develop a bar at around 6~Gyrs (as shown by $\eta$) and, according to the strength, reach their peak at 8~Gyrs and evolve secularly after the \textit{buckling} phase. The buckling is a well-known vertical instability that arises due to the on-going bar instability and is related to the vertical to radial velocity dispersion ratio, $\sigma_{z}/\sigma_{r}$ \citep{Toomre1966,Martinez-Valpuestaetal2006}. Some authors point out that violent buckling, that is, a sudden increase of $\sigma_{z}/\sigma_{r}$ during the simulation, greatly benefits the growth of bars after the buckling phase \citep[e.g.][]{Collier2020}. Here, we consider an increase of $100\%$ in the dispersion ratio from $t = 0$ to the moment the buckling is ignited, as a strong buckling~\cite[e.g.][]{Lokas2019}. From Fig.~\ref{fig:disp_curves}, we notice that $\sigma_{r}$ is always greater than $\sigma_{z}$ for models with a strong bar, hence the dispersion ratio stays the same for most of the simulation. This implies that our models do not have strong bucklings, which means that this is not the deciding factor in models that possess strong bars. However, only the models that reach the buckling instability stage, such as SGS31p or SGS34p, are able to sustain their bars, as shown by Fig.~\ref{fig:strength}. On the contrary, models with $\epsilon = 0.001$, like SGS36p, develop a bar but are unable to maintain it. This indicates that the buckling may be relevant for bars to hit their secular phase, but such study is out of the scope of this work. We aim to explain why models with small softenings cannot reach the secular phase in the next section.

\begin{table}[t]
\centering
\caption{Structural properties of the discs in runs with extra particles. \textit{Column 1}: model name. \textit{Column 2}: Toomre's criterion at $t = 0$. \textit{Column 3}: radial dispersion deviation at $t = 2.2$\,Gyrs. \textit{Column 4}: tangential dispersion deviation at $t = 2.2$\,Gyrs. \textit{Column 5}: vertical dispersion deviation at $t = 2.2$\,Gyrs. \textit{Column 6}: distorsion parameter at $t = 12$\,Gyrs. \textit{Column 7}: bar strength at $t = 12$\,Gyrs.}
\label{tab:t2}
\begin{adjustbox}{width=0.55\textwidth}
\noindent
\begin{tabular}{@{}lrcccrr}
\toprule
\textbf{Model} & $\bm{Q_{\mathrm{min}}}$ & $\bm{\overline{\sigma}_{r}}$ & $\bm{\overline{\sigma}_{\mathrm{t}}}$ & $\bm{\overline{\sigma}_{z}}$ & $\bm{\eta_{\mathrm{e}}}$ & $\bm{A_{2}}$\\ 
\multicolumn{1}{c}{(1)} & \multicolumn{1}{c}{(2)} & \multicolumn{1}{c}{(3)} & (4) & (5) & (6) & (7)\\ \midrule
SGS1p & 1.24 & 0.320 & 0.039 & 0.063 & 0.152 & 0.661\\
SGS2p & 1.24 & 0.313 & 0.044 & 0.055 & 0.122 & 0.623\\
SGS3p & 1.22 & 0.330 & 0.038 & 0.056 & 0.158 & 0.580\\
SGS4p & 1.24 & 0.355 & 0.058 & 0.055 & 0.143 & 0.654\\
SGS5p & 1.24 & 0.334 & 0.041 & 0.056 & 0.095 & 0.473\\
SGS6p & 1.24 & 0.368 & 0.078 & 0.095 & 0.008 & 0.106\\
SGS7p & 1.26 & 0.275 & 0.025 & 0.043 & 0.130 & 0.624\\
SGS8p & 1.24 & 0.288 & 0.035 & 0.039 & 0.141 & 0.685\\
SGS9p & 1.24 & 0.303 & 0.036 & 0.043 & 0.112 & 0.639\\
SGS10p & 1.24 & 0.319 & 0.044 & 0.047 & 0.140 & 0.665\\
SGS11p & 1.25 & 0.314 & 0.040 & 0.072 & 0.149 & 0.615\\
SGS12p & 1.25 & 0.390 & 0.094 & 0.063 & 0.004 & 0.050\\
SGS13p & 1.27 & 0.272 & 0.029 & 0.056 & 0.145 & 0.656\\
SGS14p & 1.26 & 0.303 & 0.037 & 0.037 & 0.115 & 0.642\\ 
SGS15p & 1.27 & 0.284 & 0.030 & 0.071 & 0.120 & 0.622\\
SGS16p & 1.28 & 0.304 & 0.034 & 0.046 & 0.004 & 0.344\\
SGS17p & 1.27 & 0.296 & 0.027 & 0.051 & 0.136 & 0.610\\
SGS18p & 1.28 & 0.316 & 0.050 & 0.043 & 0.008 & 0.065\\
SGS19p & 1.28 & 0.268 & 0.033 & 0.032 & 0.051 & 0.519\\ 
SGS20p & 1.30 & 0.284 & 0.031 & 0.033 & 0.112 & 0.606\\ 
SGS21p & 1.28 & 0.274 & 0.026 & 0.036 & 0.065 & 0.545\\
SGS22p & 1.29 & 0.289 & 0.028 & 0.041 & 0.094 & 0.662\\
SGS23p & 1.29 & 0.288 & 0.026 & 0.036 & 0.115 & 0.631\\
SGS24p & 1.30 & 0.298 & 0.042 & 0.045 & 0.023 & 0.107\\
SGS25p & 1.31 & 0.255 & 0.032 & 0.039 & 0.057 & 0.636\\ 
SGS26p & 1.28 & 0.264 & 0.032 & 0.035 & 0.068 & 0.535\\ 
SGS27p & 1.31 & 0.260 & 0.035 & 0.042 & 0.067 & 0.545\\
SGS28p & 1.30 & 0.272 & 0.032 & 0.032 & 0.081 & 0.582\\
SGS29p & 1.30 & 0.271 & 0.035 & 0.042 & 0.122 & 0.609\\
SGS30p & 1.31 & 0.274 & 0.050 & 0.033 & 0.007 & 0.139\\
SGS31p & 1.32 & 0.256 & 0.038 & 0.028 & 0.020 & 0.527\\
SGS32p & 1.31 & 0.261 & 0.033 & 0.027 & 0.022 & 0.492\\
SGS33p & 1.31 & 0.260 & 0.030 & 0.026 & 0.057 & 0.542\\
SGS34p & 1.31 & 0.262 & 0.033 & 0.039 & 0.036 & 0.610\\
SGS35p & 1.30 & 0.256 & 0.033 & 0.038 & 0.056 & 0.543\\
SGS36p & 1.31 & 0.267 & 0.047 & 0.040 & 0.007 & 0.138\\
\bottomrule
\end{tabular}
\end{adjustbox}
\end{table} 

\subsection{Vertical acceleration profiles} \label{sec:vertical-acc-profiles}

\begin{figure}
\begin{adjustwidth}{-\extralength}{0cm}
\centering
\includegraphics[scale=0.49]{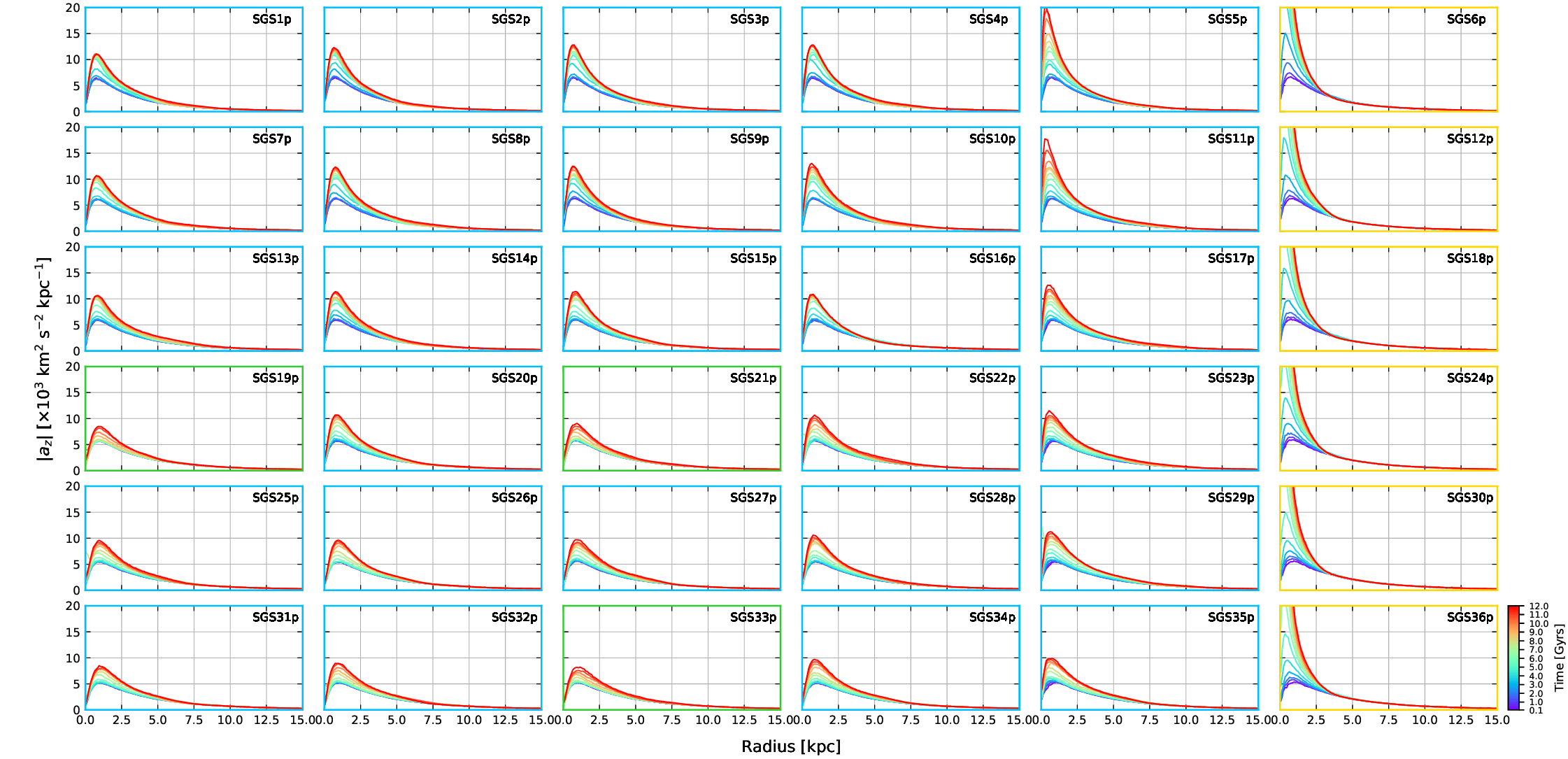}
\end{adjustwidth}
\caption{Vertical acceleration profile, $|a_{z}|$, of every discs in our high resolution models. Different colors show the temporal evolution of each simulation. The color code for the frame on each plot represents the same as in Figures~\ref{fig:snapshots} and~\ref{fig:snapshots_cont}.}
\label{fig:map-accelerations}
\end{figure}

Models with $\epsilon = 0.001$, for either low or high resolution, display an odd behaviour when compared to other models with higher softening lengths. This scenario is depicted in Fig.~\ref{fig:strength}, where SGS36p has a bar peaking at around 5~Gyrs but then it dilutes at 12~Gyrs; this phenomenon occurs in several other models with the same amount of softening, and even at a higher resolution\footnote{We ran SGS36 with $N \approx 6\times10^{6}$ and the bar is still suppressed, although it should be more physically consistent, since the errors in force calculations have smaller impact than for models with less particles.}. SGS31p ($\epsilon=0.1$) and SGS34p ($\epsilon=0.01$) develop more or less in the same manner. In fact, all high resolution models in Table~\ref{tab:t2} with $\epsilon \geq 0.005$ have similar dynamics. 

To shed some light on the disparity between the evolution of these models and how it could relate to $\epsilon$, we track the vertical acceleration profiles, $|a_{z}|$, of every particle in our high resolution set and map the results for different times (Fig.~\ref{fig:map-accelerations}). Each plot in the grid shows the evolution of $|a_{z}|$ against $r$ for every model in Table~\ref{tab:t2}. It is clear that vertical accelerations in the centre of discs increase with time, independently of the bar's properties. Accelerations increase in a lesser rate for models with higher $m_{\mathrm{d}}$. Decreasing $\epsilon$ appears to have a positive impact on the central acceleration rate; both these effects are tightly linked to the bar strength. For instance, SGS19 is less accelerated than SGS25 despite having less disc mass. Conversely, SGS20 is more accelerated than SGS21 despite having a higher softening. Both SGS19 and SGS21 have noticeably weaker bars than most of the models in Table~\ref{tab:t2}, at least according to their $\eta_{\mathrm{e}}$ values. Nevertheless, if bar strength remains comparable, which happens for the first two rows in Fig.~\ref{fig:map-accelerations}, we see that accelerations increase when $\epsilon$ decreases. An increase in vertical acceleration is not always correlated to bar strength. For example, SGS5 has a comparably weaker bar than SGS3 or SGS4 (following both $\eta_{\mathrm{e}}$ and $A_{2}$), despite being highly accelerated. This is where the noise introduced by an excess in softening starts to dominate the acceleration profile, overriding any effect induced by the bar; such event occurs for models with $\epsilon \leq 0.005$, although it diminishes when $m_{\mathrm{d}}$ increases. Simulations with $\epsilon = 0.001$ appear to be physically unreliable since their vertical acceleration profiles reach values that do not correspond to a disc-like system. For simulations with this softening length, the particles with the highest accelerations are concentrated in the centre, which could explain their dynamics, particularly why bars are suppressed in these models.

We now summarize the main results emerging from Fig.~\ref{fig:map-accelerations}. Any trend involving accelerations depends on two effects: numerical noise related to $\epsilon$ and the bar instability. At first glance, it is hard to distinguish which effect is the one meddling with the accelerations. For example, the strength of bars shown by  $\eta_{\mathrm{e}}$ in Table~\ref{tab:t2} is comparively similar between models SGS1p, SGS2p, SGS3p, SGS4p and SGS5p, but the $|a_{z}|$ peak is definitely greater for SGS4p and SGS5p, which have lower softening values. Conversely, models with the weakest bars at $t = 12$~Gyrs, e.g. SGS19p, SGS21p or SGS31p, are significantly less accelerated than models with stronger bars. 
However, we are able to say that lower softenings accelerate the central regions of phase-space, which may contribute to the bar's supression or destruction; such effect is much less prevalent as one increases $\epsilon$.

Although our intent is to make our study as comprehensive as possible, there are still many scenarios left to explore. Our range for $m_{\mathrm{d}}$ is limited and, although it is assumed that higher $m_{\mathrm{d}}$ values should generate more bar-stable models, it would be interesting to study its interaction with $\epsilon$. The models are also relatively simple, containing only two dynamical components and one profile for each component; gas components and bulges not only intervene in bar dynamics, they also must be modeled using some softening recipe and it has been shown that the mass of a gaseous component and its softening length value may affect how a bar is triggered and formed~\citep{Villa-VargasShlosmanHeller2010}. 

We also used the same fixed $\epsilon$ for both components regardless of $N$, but there are other strategies available (see appendix~\ref{sec:apen1}) to set an $\epsilon$ for each component. One may use an adaptive approach to set the softening lengths, but such approach still requires to set upper and lower limits for each species.

%

\section{Conclusions} \label{sec:conclusions}

We ran a series of numerical models to follow the influence of two parameters ($\epsilon$ and $m_{\mathrm{d}}$) on bar formation and secular evolution, also accounting for the effect of $N$ in particle dynamics. We choose $0.001 \leq \epsilon \leq 0.1$ as our softening window, which sets it around the values opted by the majority of numerical studies, including ones that use an adaptative algorithm. The disc mass range is adjusted using the standard disc formation theory and is set to $0.035 \leq m_{\mathrm{d}} \leq 0.05$, which properly fits observed disc galaxies. The results of these experiments are summarized as follows.

\begin{itemize}
\item[$\diamond$] The set of galaxy models considered here vary in mass and disc extension but have, by construction, nearly the same initial value of $Q_{\mathrm{min}}$ (${\approx} 1.1$). In this sense, the models are marginally stable against bar formation.
\item[$\diamond$] Further evolution of the Toomre's criterion does not reflect bar instability for all of our models, particularly for models with the lowest softening values, i.e. $\epsilon \leq 0.005$~kpc (see section~\ref{sec:simulations}). This behaviour still occurs after increasing the number of particles of the models. 
\item[$\diamond$] We find that $\eta$ yields an accurate picture of the bar's formation and evolution (see Fig.\,\,\ref{fig:eta_curves}). We confirm that a threshold of $\eta_{\mathrm{th}} = 0.02$ effectively indicates recent bar formation. Additionally, this parameter allows us to observe the formation, destruction and resurgence of the bar, which is not possible with other indicators such as the bar strength, $A_2$. The distortion parameter is more apt at measuring when the bar has formed and A2 is more apt at identifying the different phases of bar evolution (rapid growth, buckling, secular growth). In this case, we are inclined to use the distortion parameter because we are interested in when the bar appears instead of determining how strong it might get.
\item[$\diamond$] We find a close relationship between $\epsilon$ and $m_{\mathrm{d}}$ that affects how fast the bar forms, its strength and length. We notice that the empirical linear dependencies, that is, $\eta \propto m_{\mathrm{d}}^{-1}$ and $\eta \propto \epsilon$, between our simulations and ($\epsilon, m_{\mathrm{d}}$) pairs do not hold when these two parameters interact. Thus, the process of choosing an appropriate softening value seems to be more complicated than previously thought.
\item[$\diamond$] The vertical acceleration profile is a better estimator of the disc heating than the velocity dispersions. Our models with small softening values ($\epsilon \leq 0.005$~kpc) tend to accumulate and accelerate particles at the centre of discs, driving them toward chaotic motion. Such effect normally results in the destruction of the bar.
\item[$\diamond$] While dispersions do increase at the centre of unbarred models, such increase is not that extreme and may be explained by natural causes (e.g. disc mass redistribution, bar residue, non-axisymmetric distortions, etc.). Given that the central disc density increases with time in our models, regardless of parameter values, it is also reasonable to assume that the near circular particle orbits have simply shrunk, dimly increasing the dispersions. However, none of these effects necessarily explain the bar dilution. Because accelerations are not affected by these natural causes, they are better tracers of these unphysical/odd behaviors.
\item[$\diamond$] We are also able to conclude that, for particle resolutions close to $N \approx 10^{7}$, softening values lower than 0.005~kpc are not well-suited to reproduce the bar instability. However, this depends on the disc mass fraction. If $m_{\mathrm{d}} \simeq 0.05$, the bar appears to cohabit with numerical noise introduced by small softening values, specifically, $\epsilon = 0.005$~kpc. Models with lower $m_{\mathrm{d}}$ values ($\lesssim 0.040$) are not able to overcome the interference of small softening values, affecting the behaviour of the bar (see Fig.~\ref{fig:map-accelerations}).
\end{itemize}


Applying a comprehensive set of numerical tools to study the bars in collisionless disc models, we conclude that such analysis is necessary to understand the effects induced by $\epsilon$ and its interplay with $m_{\mathrm{d}}$ and $N$. Our analysis indicates that the vertical acceleration profile, $|a_z|$, adequately describes the behaviour of our disc, particularly, why there is bar supression in some of our models. However, since both the bar and numerical noise related to $\epsilon$ heat up the vertical acceleration profile, it is not trivial to tell which one is affecting the acceleration at any particular region. Numerical noise usually gathers at the inner-most regions of the disc, but it is unclear at what point in phase-space the bar starts to affect these regions too. This means that there is a close relationship between the bar distortion and $\epsilon$, one that is much more intricate than previously thought~\citep[cf.][]{IannuzziAthanassoula2013}. Balance amid these effects may result in more faithful simulations of galactic discs.

\vspace{6pt} 





\authorcontributions{
Conceptualization, AL, RG; 
methodology, AL, RG, IFC;
software, AL, RG;
validation, AL;
formal analysis, AL, RG;
resources, AL, IFC;
data curation, AL, RG;
writing---original draft preparation, AL;
writing---review and editing, RG, IFC;
visualization, AL, RG, IFC;
supervision, RG, IFC;
project administration, RG, IFC;
funding acquisition, IFC.
All authors have read and agreed to the published version of the manuscript.
}

\funding{AL thanks and acknowledges CONAHCyT for the doctoral grant and CONAHCyT project CDF2019. IFC acknowledges IPN-SIP grant 20241455.}

\dataavailability{All the simulations included in this manuscript can be shared upon reasonable request to the corresponding author.} 

\acknowledgments{The authors acknowledge UNAM (\textit{Universidad Nacional Aut\'onoma de M\'exico}) for the HPC resources of Abacus and At\'ocatl supercomputers, which have contributed to the research results reported within this manuscript.}

\conflictsofinterest{The authors declare no conflicts of interest. The funders had no role in the design of the study; in the collection, analyses, or interpretation of data; in the writing of the manuscript; or in the decision to publish the results.}

\appendixtitles{no} 
\appendixstart
\appendix

\section[\appendixname~\thesection]{The effect of unequal softening lengths} \label{sec:apen1}

\begin{figure}[t]
\centering
\includegraphics[scale=0.8]{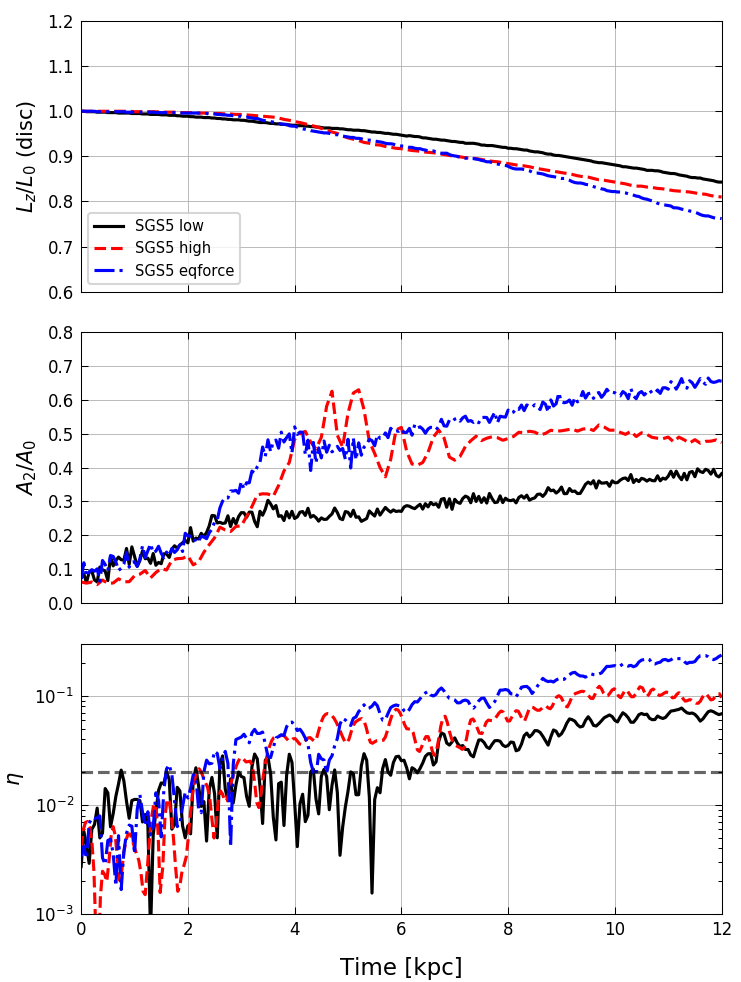}
\caption{Properties of discs for SGS5 (black full line), SGS5p (red dashed line) and SGS5eq (blue dash-dotted line). \textit{Top}: normalized angular momentum transfer, $L_z/L_0$. \textit{Middle}: normalized $m=2$ Fourier amplitude of the surface density. \textit{Bottom}: distorsion parameter, $\eta$. The horizontal dashed line is the bar threshold given by $\eta_{\mathrm{th}} = 0.02$.}
\label{fig:ang-mom-a2-apen}
\end{figure}

\begin{figure}[t]
\begin{adjustwidth}{-\extralength}{0cm}
\includegraphics[scale=0.73]{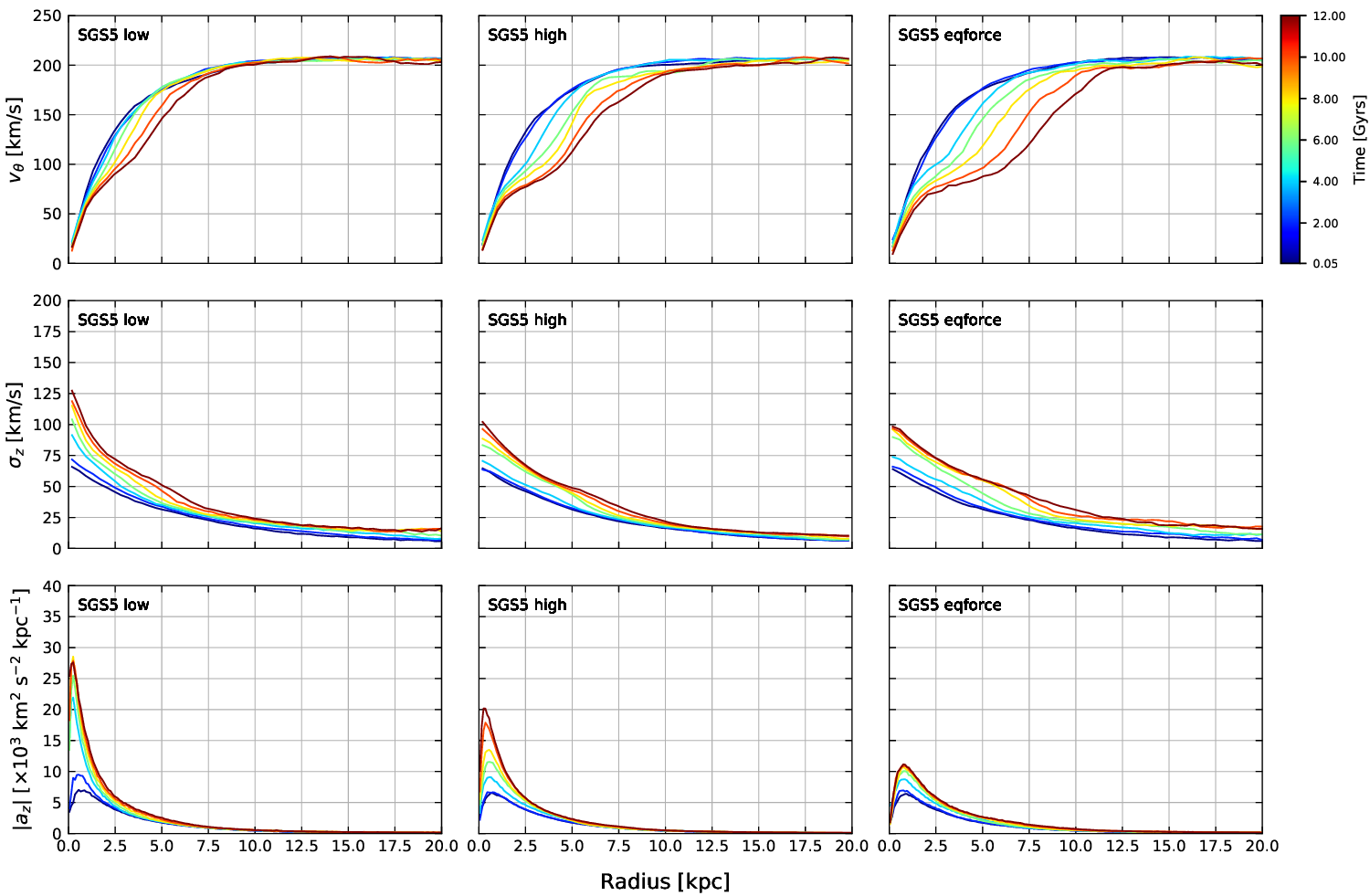}
\end{adjustwidth}
\caption{Evolution of the tangential velocities (top row), vertical velocity dispersion (center row) and vertical acceleration profile (bottom row) of models SGS5, SGS5p and SGSeq. The color code for the frame on each plot represents the same as in Figures~\ref{fig:snapshots} and~\ref{fig:snapshots_cont}.}
\label{fig:rc_sigma_z_acc_apen}
\end{figure}

In order to avoid using the same particles masses for halo and disc components, a usual practice is to assign different softening lengths to each one. Usually, the softening length of one component is chosen in terms of the other to guarantee the same gravitational force between particles of different species~\citep{McmillanDehnen2007}. This way one avoids undesired effects such as two-body scattering, and gravitational drag force. 

In this appendix, we repeat simulation SGS5 in Table~\ref{tab:t1}, this time applying the softening recipe described above. So, the softening length of halo particles is set to $\epsilon_{\mathrm{h}}=0.041$, while keeping $\epsilon_{\mathrm{d}}=0.005$ for the disc particles. The initial conditions were created with $\epsilon=0.041$. As mentioned in section~\ref{sec:ics},  \textsc{GALIC} uses a single value for $\epsilon$ for all components while producing its equilibrium galaxy models. For comparison, an equivalent NFW halo with $N=10^6$ that minimizes the acceleration errors will have $\epsilon=0.288$~\citep{Zhan2006}.

The outcome of this run, named SGS5eq, is now compared to the original low resolution SGS5 model and the high resolution SGS5p rerun. Fig.~\ref{fig:ang-mom-a2-apen} shows the angular momentum transfer (top), bar amplitude (middle) and distorsion paramenter (bottom) for these three models. The upper panel shows that all three runs behave similarly until their respective bars emerge at $t\sim 2.5$\,Gyrs. After that, the discs lose their momentum proportionally to the bar strength. From the middle panel, we notice that SGS5 and SGS5eq start in a similar manner (up to ${\sim}2.5$\,Gyrs), but there is a clear departure from this point where the bar of SGS5eq rapidly grows until the end of the simulation. The evolution of the bar in SGS5p is similar to SGS5eq but there is a small delay in its growth. The buckling phase in SGS5p (${\sim}5$\,Gyrs) is visibly stronger than both SGS5 and SGS5eq, but the bar growth in SGS5p stagnates after this phase, contrary to SGS5eq, where the bar continues to grow after the end of the buckling. The behaviour of the bar strength may also be observed in the tranfer of angular momentum, where the strongest bar induces a higher transfer rate. The distorsion parameter graphs reflect the delays suffered by the bars in SGS5p and, most notibly, SGS5.

Furthermore, by comparing the velocity dispersions profiles and the tangential velocities in Fig.~\ref{fig:rc_sigma_z_acc_apen}, it can be seen that in the original SGS5 run there is an excess in the velocity dispersion at the centre of the disc at late stages of bar formation, and more intense suppression of the tangential velocity in SGS5eq due to a stronger bar. In this sense, the tangential velocity and vertical velocity dispersion of model SGS5p lies between the SGS5 and SGS5eq. One key difference in SGS5eq is that the vertical heating ($\sigma_z$) is not confined to the region occupied by the bar, having enhanced dispersions for the whole disc. This fact would not be expected as heating should be inhibited by setting different $\epsilon$ values for each galaxy component. We suggest that it is rather due to the bar dynamics itself, that favours the transfer of orbital momentum into the vertical motions (see top panel in Fig.~\ref{fig:ang-mom-a2-apen}).

Finally, we address the evolution of the disc particles vertical accelerations shown in the bottom of Fig.~\ref{fig:rc_sigma_z_acc_apen}. Up to $t=0.1$ Gyrs, the discs in all three models conserve the same dynamical equilibrium despite the value of $\epsilon$. After $t\sim2$ Gyrs the original SGS5 model rapidly heats up in the centre albeit having a weaker bar. We attribute this difference to the choice of the single $\epsilon$ value. In fact, the disparity between SGS5 and SGS5eq in terms of bar strength also happens at $t\sim2$, confirming that enhanced accelerations are, in the experiments shown here, the culprit for bar supression. Again, the high resolution model shows an intermediate behaviour.

The intermediate behaviour of SGS5p falls right between SGS5 and SGS5eq due to its smaller mean interparticle distance to $\epsilon$ ratio, which puts it closer to the ratio of SGS5eq, given that the latter has a larger $\epsilon$ for the halo component.

The results in this section imply that the choice of the softening value has an impact on long term bar evolution, rather than any sudden change of dynamics at the beginning due to the difference in softening values between the ICs and the actual run.

\begin{adjustwidth}{-\extralength}{0cm}

\reftitle{References}


\bibliography{Definitions/biblio}

\PublishersNote{}
\end{adjustwidth}
\end{document}